\documentclass[showpacs,preprintnumbers,amsmath,amssymb,prd]{revtex4}
\usepackage{graphicx}
\usepackage{amsmath,amssymb}

\begin{document}

\title{Nambu-Jona Lasinio and Nonlinear Sigma Models in Condensed Matter Systems} 

\author{Ryosuke Yoshii}
\affiliation{Department of Physics, Chuo University, 1-13-27 Kasuga, Bunkyo-ku, Tokyo 112-8551, Japan} 
\affiliation{Research and Education Center for Natural Sciences, Keio University, 4-1-1 Hiyoshi, Kanagawa 223-8521, Japan}

\author{Muneto Nitta}
\affiliation{Department of Physics,  Keio University, 4-1-1 Hiyoshi, Kanagawa 223-8521, Japan}
\affiliation{Research and Education Center for Natural Sciences, Keio University, 4-1-1 Hiyoshi, Kanagawa 223-8521, Japan}

\date{\today}
\begin{abstract}
We review various connections between condensed matter systems with the Nambu-Jona Lasinio model and nonlinear sigma models. 
The field theoretical description of interacting systems offers a systematic framework to describe the dynamical generation of condensates. 
Resent findings of a duality between the Nambu-Jona Lasinio model and the nonlinear sigma model enables us 
to investigate various properties underlying both theories. 
In this review we mainly focus on inhomogeneous condensations in static situations. 
The various methods developed in the Nambu-Jona Lasinio model reveal the inhomogeneous phase structures and 
also yield new inhomogeneous solutions in the nonlinear sigma model owing to the duality. 
The recent progress on interacting systems in finite systems is also reviewed. 
\end{abstract}


\maketitle

\section{Introduction}
Fermionic systems with many body interactions affect the ground state structure. 
In the case of repulsive interaction, the vacuum state and the low energy excited states are well described by the Fermi liquid theory in which the notions of adiabatic continuity and quasi-particles play important roles \cite{Landau,Nozieres}. 
In the framework, the low energy phenomena can be explained by the dressed particle and a similar behavior with the free fermionic system is obtained. 
This approach succeeds to explain transport phenomena in normal metals; the electron's behavior in metals has been known to be similar one of a free particle though the microscopic model involves manybody effects. 
On the other hand, the case of attractive interaction is highly nontrivial. 
In that case, the vacuum structure is not continuously describable from a noninteracting system in the sense of adiabatic continuity; rather there is a phase transition. 
In other words, phenomena taking place in such system are non-perturbative. 

In condensed matter systems, such phenomena have been found in the context of superconductivity. 
Electrons have Coulomb repulsions between all pairs, however a lattice system can result in the attractive interaction mediated by a phonon. 
The pioneering work done by Cooper treats a two-body problem in addition to the background Fermi sea (conduction band). 
It is shown that the attractive interaction, even though the strength is infinitesimally weak, results in the formation of a bound state below the characteristic temperature. 
This means that the Fermi sea is unstable against the attractive interactions (Cooper instability) \cite{Cooper}.
The succeeding work done by Bardeen-Cooper-Schrieffer (BCS) revealed a mechanism of the superconductivity \cite{BCS} 
and results drown by the model reproduce those of a phenomenological model, so-called the Ginzburg-Landau model, proposed before \cite{Ginzburg} which describes the experimental results quite well \cite{Gorkov}. 
In their paper, the vacuum state in the presence of the attractive interaction is represented by a state consisting of the superposition of states with different particle numbers. 
This is a consequence of the breaking of the U(1) symmetry of the wave function.

In the high energy physics, the BCS scenario has been reinterpreted as a spontaneous symmetry breaking by Nambu and Jona-Lasinio (NJL) \cite{NJL}. 
In their model, the chiral symmetry is spontaneously broken and the mass of fermions, which is zero in the unbroken phase, is dynamically generated. 
The idea has led to the other important notions such as Nambu-Goldstone (NG) bosons \cite{Nambu,Ambegaokar,Goldstone,GSW} and the Higgs mechanism  \cite{Anderson,Kibble,EB,Higgs} in the presence of a coupling with gauge fields. 
After the pioneering works, the notion of the spontaneous symmetry breaking is widely used in the various fields of physics. 
Later a similar model in 1+1 dimensions, so called the (chiral) Gross-Neveu (GN) model, has been investigated \cite{GN}. 
The model itself is almost the same with the one used in Ref.~\cite{NJL}, however the model holds the renormalizability owing to low dimensionality. 

Recently, a connection between the NJL model and nonlinear $\sigma$ (NL$\sigma$) models is recognized \cite{footnote1}. 
The NL$\sigma$ model describes free bosons with a constraint resulting in a nonlinear derivative interaction among bosons.
The model was originally introduced as a toy model of QCD; the both theories share many non-perturbation properties such as asymptotic freedom, dynamical mass generation and instantons. 
Recently one of NL$\sigma$ models known as the $\mathbb{C}P^{N-1}$ model attracts renewed interests 
since the model is rederived as the effective model of a vortex line in a non-Abelian gauge theory \cite{Hanany,Auzzi,Eto,Tong,Eto2,Shifman}. 
In condensed matter systems the O($N$) sigma model naturally arises from the continuum limit of the Heisenberg model which describes  spins on a lattice with the spin exchange interactions \cite{Haldane, Affleck}. 
The quantum phase transition (deconfined criticality) is proposed in antiferromagnetic systems \cite{Senthil, Sudbo}. 
The NL$\sigma$ model with a topological term is known to describe the integer quantum Hall effect \cite{Affleck, Pruisken}. 
The NL$\sigma$ model especially the $\mathbb{C}P^{N-1}$ sigma model shares deep roots with the NJL model through supersymmetry \cite{SUSYCPN}. 
A similar structure bertween partition functions of the two theories in 2+1 dimension is also found \cite{Filothodoros}. 
Another observation reveals a clear correspondence between the two theories in 1+1 dimensions in a microscopic level \cite{YoshiiNitta17}.

The above topics are restrictive. 
For instance, the presence of a magnetic impurity in a metal induces the Kondo effect in which the confinement and asymptotic freedom of the spin on the impurity takes places depending on the energy scale \cite{Kondo,Hewson}. 
In the presence of a strong magnetic field, fractional particles consisting of magnetic fluxes and electrons condensate. 
Consequently, the quantization of the conductance takes place (Hall effect), whose effective theory is described by Chern-Simons model \cite{Zhang,Zee}. 
Although there are huge area of the cross section between condensed matter physics and high energy physics, we leave those issues for the other articles and focus on topics about systems described by the NJL model. 

This article is organized as follows. 
In Sec.\ \ref{NJLandCondMatt}, we briefly review the BCS model and the equivalence between the BCS model and NJL model in suitable approximations. 
We also show the discretized model which describes superconducting or superfluid phenomena in materials or cold atom systems. 
As another important example, we also introduce a polyacetylene system. 
In Sec.\ \ref{InhomogeneousSols}, we discuss inhomogeneous condensates. 
Inhomogeneous solutions and their properties are explained there. 
Theoretical structures which enable us to find inhomogeneous solutions are presented. 
In Sec.\ \ref{NJLandNS}, a connection between the NL$\sigma$ model and NJL model is briefly introduced. 
The various solutions obtained from the knowledge of the NJL model are discussed. 
Sec.\ 5 is devoted to a summary and discussion.

\section{NJL model and condensed matter systems}{\label{NJLandCondMatt}} 
Throughout this article, we will concentrate on 1+1 dimensional systems in the condensed matter except for Sec.\ \ref{3DNJLNLS}. 
The remarkable features often arise by focusing on the 1+1 dimensional systems, 
e.g.~the conformal invariance, integrable structure, bosonization or fermionaization, and so on. 
This simplification does not completely spoil the validity of the following results in the higher dimensions at least in the qualitative level. 
Also, the progress on the experimental techniques enable us to realize low dimensional systems in laboratories. 
For instance, the carbon nanotube is well described by a quasi one dimensional model. 
In the system, the chirality and spin degrees of freedoms form internal SU(4) symmetry and the theoretically proposed phenomena have been observed experimentally \cite{Jarillo}. 
More recently, the laser technique opens a new area of research, so called ultracold atomic physics, in which $\sim$ 1000 atoms can be trapped in an egg carton shape potential and one can design the shape of an atomic gas almost arbitrary. 
The quasi one dimensional system (cigar shape trap) is also realized in cold atoms and it is again well described by a one dimensional model \cite{Liao}. 
A ring shape trap is also realized experimentally and the persistent current along the ring is reported \cite{Ryu}. 
The conducting polymer, e.g., polyacetylene, is also a quasi one dimensional system \cite{Heeger}. 
 
In the following we always refer the high energy counterparts as the NJL model without specifying the details for convenience 
\cite{footnote2}.

\subsection{Interacting fermion}
In this subsection we introduce the BCS model and briefly show the appearance of the superconducting gap in the infinite system. 
We start from the following simple model: 
\begin{align}
H=&\sum_{\sigma}\int dx \psi_{\sigma}^\dagger(x) \left(-\frac{\hbar^2}{2m}\partial_x^2-E_{\mathrm{F}}\right)\psi_{\sigma}(x)
+\int dx dx^\prime \psi_{\uparrow}^\dagger(x)\psi_{\downarrow}^\dagger(x^\prime) V(x-x^\prime)\psi_{\downarrow}(x^\prime)\psi_{\uparrow}(x)\nonumber\\
=&\sum_{\sigma}\int dx \psi_{\sigma}^\dagger(x) \left(-\frac{\hbar^2}{2m}\partial_x^2-E_{\mathrm{F}}\right)\psi_{\sigma}(x)
-g\int dx \psi_{\uparrow}^\dagger(x)\psi_{\downarrow}^\dagger(x) \psi_{\downarrow}(x)\psi_{\uparrow}(x),
\label{IntFer}
\end{align}
where $\psi_{\sigma}$ stands for the wave function of spin $\sigma$ fermions and $V(x-x^\prime)$ is the interaction between two fermions and $E_{\mathrm{F}}$ is the Fermi energy. 
For simplicity, we focus on the hardcore attractive interaction $V=-g\delta(x-x^\prime)$ which mimics the short range attractive interaction, for example, mediated by the phonon. 
It is known that two electrons put on the Fermi surface can always form the bound state in the presence of the attractive interaction, even though the interaction is infinitesimally small. 
This phenomenon shows that the Fermi surface is unstable against the attractive interaction between fermions and is called the Cooper instability \cite{Cooper}. 
In this argument the Fermi surface is treated as the background field and the problem itself is almost two-body problem, but the binding energy is found to be non-perturbative from the following reason. 
The binding energy can be calculated by solving the two-body problem and is found to be proportional to $\exp(-1/2g\rho)$ with the density of state $\rho$. 
This result shows that the perturbative treatment is no longer valid to explain this phenomenon since small $g$ expansion of $\exp(-1/2g\rho)$ is not possible. 
The bound state is called a Cooper pair and the superconducting phenomenon is understood as the Cooper pair condensation where the Fermi sea is no longer the background field but the occupied state of Cooper pairs. 
This observation leads to the identification of the coherent state $\prod_k A_k\exp(\alpha_k c^\dagger_{k,\uparrow}c^\dagger_{-k,\downarrow})|0\rangle=\prod_k \left(A_k+A_k \alpha_k c^\dagger_{k,\uparrow}c^\dagger_{-k,\downarrow}\right)|0\rangle$ with the ground state wavefunction \cite{footnote3}.  
In the BCS paper, the coherent state is considered as a trial function and the parameters $\alpha_k$ and $A_k$ are determined from the variational principle. 

In order to explore the ground state of the system, here we employ the mean field approximation which corresponds to the large-$N$ approximation in the NJL context. 
According to the above argument about the coherent state, the ground state should have the non-vanishing vacuum expectation value (VEV) of the Cooper pair creation operator $c^\dagger_{k, \uparrow} c^\dagger_{-k, \downarrow}$. 
Thus, we employ the following mean field approximation for the interaction term 
\begin{equation}
-g\int dx \psi_{\uparrow}^\dagger(x)\psi_{\downarrow}^\dagger(x) \psi_{\downarrow}(x)\psi_{\uparrow}(x)
\simeq \int dx \Delta(x) \psi_{\uparrow}^\dagger(x)\psi_{\downarrow}^\dagger(x) +\int dx \Delta^\ast (x) \psi_{\downarrow}(x)\psi_{\uparrow}(x) 
+\frac{1}{g}\int dx |\Delta(x)|^2,
\label{meanfield}
\end{equation}
where we have introduced the pair potential $\Delta(x)$ as 
\begin{equation}
\Delta\equiv -g\langle \psi_{\downarrow}(x)\psi_{\uparrow}(x)\rangle
=g\langle \psi_{\uparrow}(x)\psi_{\downarrow}(x)\rangle. 
\label{gapeq}
\end{equation}
Here the average $\langle \cdots\rangle$ above stands for the average with respect to the ground state. 

\begin{figure}
\includegraphics[width=14 cm]{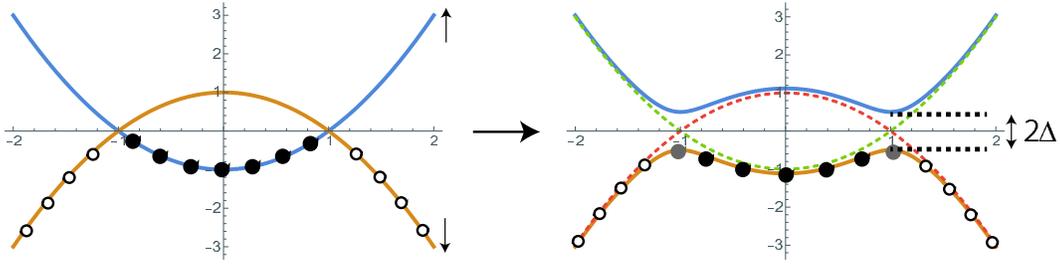}
\caption{The energy spectrum for the normal state (left figure) and the superconducting state (right figure). 
In the left figure the electron band (downward convex line) and the hole representation for the band (convex upward line) are shown. 
In the terminology of the electron band, the states under the Fermi level are filled (black circles) while the states above the Fermi level are filled (white circles) for the hole terminology. 
In the presence of the superconducting correlation, the gap in the energy spectrum appears (right figure).}
\label{figBCS}
\end{figure}

In the following, we omit the $\int dx |\Delta|^2/g$ term for a while. 
The Hamiltonian (\ref{IntFer}) with the mean field approximation (\ref{meanfield}) can be rewritten in the following matrix form. 
\begin{equation}
H=
\int dx\left(
\psi_{\uparrow}^\dagger\ \psi_{\downarrow} 
\right)
\left(
\begin{array}{cc}
-\hbar^2\partial_x^2/2m-E_\mathrm{F} & \Delta \\
\Delta^\ast & \hbar^2\partial_x^2/2m+E_\mathrm{F}
\end{array}
\right)
\left(
\begin{array}{cc}
\psi_{\uparrow} \\
\psi_{\downarrow}^\dagger
\end{array}
\right).
\label{NambuRep}
\end{equation}
This two component spinor representation is called the Nambu-representation and the energy spectrum of the Hamiltonian is 
given by the following equation: 
\begin{equation}
\left(
\begin{array}{cc}
-\hbar^2\partial_x^2/2m-E_\mathrm{F} & \Delta \\
\Delta^\ast & \hbar^2\partial_x^2/2m+E_\mathrm{F}
\end{array}
\right)
\left(
\begin{array}{cc}
\psi_{\uparrow} \\
\psi_{\downarrow}^\dagger
\end{array}
\right)
=E\left(
\begin{array}{cc}
\psi_{\uparrow} \\
\psi_{\downarrow}^\dagger
\end{array}
\right).
\label{BdG}
\end{equation}
This equation is called the Bogoliubov - de Gennes (BdG) equation \cite{BdG}. 
By solving Eqs.\ (\ref{BdG}) and (\ref{gapeq}) in a self-consistent manner, one can obtain the ground state and the corresponding pair potential $\Delta$. 
For a constant $\Delta$ (homogeneous condensation), the energy spectrum is easily found by the Fourier transformation 
\begin{equation}
E_k=\pm \sqrt{(\epsilon_k^0)^2+|\Delta|^2}, 
\end{equation}
where $\epsilon_k^0=\hbar^2 k^2/2m-E_\mathrm{F}$.
This energy spectrum represents the appearance of the energy gap (Fig.~\ref{figBCS}). 
In the case of zero temperature, the ground state is given by filling the all negative energy states and thus the ground state energy is given as 
\begin{equation}
E_g=-\int \frac{dk}{2\pi} \sqrt{(\epsilon_k^0)^2+|\Delta|^2}. 
\end{equation}
Using the Feynmann-Helmann theorem, one can calculate the pair potential as 
\begin{equation}
\Delta=-g\langle \psi_{\downarrow}(x)\psi_{\uparrow}(x)\rangle=-g\left\langle\frac{\delta H}{\delta \Delta^\ast}\right\rangle
=-g\frac{\delta E_g}{\delta \Delta^\ast}=g\int \frac{dk}{2\pi} \frac{\Delta}{2\sqrt{(\epsilon_k^0)^2+|\Delta|^2}}. 
\label{gapeq2}
\end{equation}
This equation is called the gap equation since it determines the amplitude of the superconducting gap. 
The same equation can also be obtained by minimizing the total energy $E_{\mathrm{tot}}=E_g+\int dx |\Delta|^2/g $ with respect to $\Delta$.

\begin{figure}
\includegraphics[width=12cm]{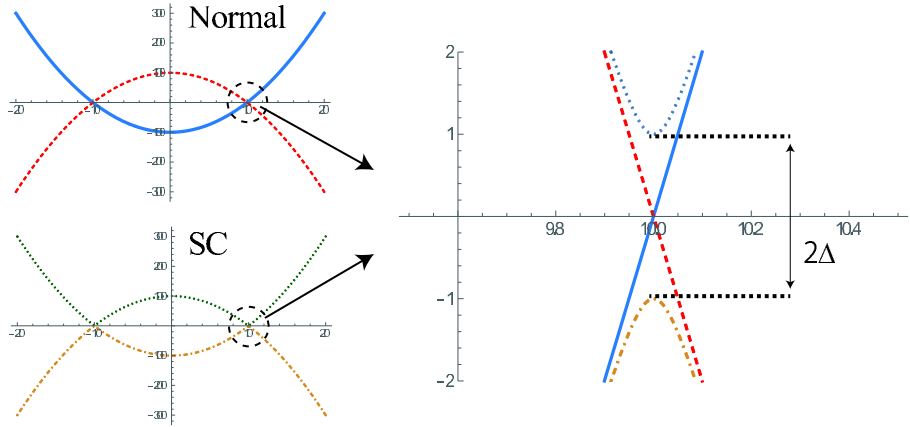}
\caption{The energy spectrum in realistic parameters. 
The top-left figure shows the energy spectrum for the normal state. 
The solid and the dashed lines describe the particle and hole energy dispersions, respectively. 
The bottom left figure shows that for the superconducting phase. 
The dotted and line-dotted lines denote the positive energy and negative energy branches, respectively.  
The superconducting gap is much smaller than the band width and the linear approximation is justified in this case (right figure).}
\label{figAndreev}
\end{figure}

In a realistic parameter setup, the pair potential (or the superconducting gap) $\Delta$ is typically $\sim$ 100 times smaller than the Fermi energy $E_{\mathrm{F}}$. 
Since only electrons in the energy window $[E_{\mathrm{F}}-\Delta, E_{\mathrm{F}}+\Delta]$ participate in the superconducting phenomena, 
which do not depend on the detailed energy structure far from the Fermi energy. 
Thus one can linearize the energy spectrum for the description of the small gap superconductors (Fig.~\ref{figAndreev}) as 
\begin{equation}
\frac{\hbar^2 k^2}{2m}-E_{\mathrm{F}} 
= \frac{\hbar^2\left[(k-k_{\mathrm{F}})+k_{\mathrm{F}}\right]^2}{2m}-E_{\mathrm{F}} 
= \frac{\hbar^2k_{\mathrm{F}}}{m}(k-k_{\mathrm{F}}) +O(k-k_{\mathrm{F}})^2. 
\end{equation}
Here we have used $E_{\mathrm{F}}=\hbar^2k_{\mathrm{F}}^2/2m$. 
The coefficient of $k-k_\mathrm{F}$ can be rewritten as $\hbar^2 k_{\mathrm{F}}/m=\hbar p_{\mathrm{F}}/m=\hbar v_{\mathrm{F}}$, 
where $p_{\mathrm{F}}$ and $v_{\mathrm{F}}$ are the Fermi momentum and the Fermi velocity, respectively. 
This approximation is called the quasi-classical or Andreev approximation \cite{Andreev}. 
The approximation is valid if $\hbar v_{\mathrm{F}} (k-k_{\mathrm{F}}) \gg \hbar^2(k-k_{\mathrm{F}})^2/m$ in the energy window $-\Delta<\hbar v_{\mathrm{F}} (k-k_{\mathrm{F}})<\Delta$. 
Thus, he criterion for the validity of the linearization is found  to be $1 \gg \Delta/mv_{\mathrm{F}}^2$ which can also be written as $k_{\mathrm{F}}\gg \Delta/\hbar v_{\mathrm{F}}$. 
This condition is equivalent to $ \hbar v_{\mathrm{F}}/\Delta=\xi \gg \ell_{\mathrm{F}}$, where $\xi$ is characteristic length of the superconductor, so called the coherent length, and $\ell_{\mathrm{F}}\sim k_{\mathrm{F}}^{-1}$ is the Fermi length. 
The length scale $\xi$ is of the order of the size of the Cooper pair and 
the length scale $\ell_{\mathrm{F}}$ is of the order of the average distance between the electrons. 
Thus, the strong interaction and/or the high density alters this situation. 
In the same time, the BCS treatment itself is no longer valid (for such systems, see Refs.\ \cite{Leggett, NSR}, 
in which the change of the chemical potential is taken into account). 

Now we reach at the following Hamiltonian: 
\begin{equation}
H=
\int dx\left(
\psi_{\uparrow}^\dagger\ \psi_{\downarrow} 
\right)
\left(
\begin{array}{cc}
-i\hbar v_{\mathrm{F}}\partial_x & \Delta \\
\Delta^\ast & i\hbar v_{\mathrm{F}}\partial_x
\end{array}
\right)
\left(
\begin{array}{cc}
\psi_{\uparrow} \\
\psi_{\downarrow}^\dagger
\end{array}
\right).
\label{NambuRep2}
\end{equation}
This model is completely the same with the one dimensional NJL model. 
It should be noted that the role of the speed of light in high energy physics is replaced by the Fermi velocity in the condensed matter physics. 

The system in which the linearized approximation is reasonable, we can set the density of states to be constant, since 
\begin{equation}
\rho=\frac{1}{2\pi}\frac{dk}{d\epsilon_k}=\frac{1}{2\pi} \frac{1}{\hbar v_{\mathrm{F}}}.
\end{equation}
Thus, one can easily solve the gap equation (\ref{gapeq}) for the homogeneous case $\Delta=\mathrm{const.}$\ with the UV cutoff $\Lambda$ as follows 
\begin{equation}
1=g\rho \int_0^\Lambda d\epsilon  \frac{1}{2\sqrt{\epsilon^2+|\Delta|^2}}=\frac{g\rho}{2} \sinh^{-1} \frac{\Lambda}{|\Delta|}. 
\end{equation}
For the weak coupling $g\rho\ll 1$, one can obtain 
\begin{equation}
|\Delta|\sim 2\Lambda \exp\left(-\frac{2}{g\rho}\right). 
\end{equation}
Here as we already mentioned in the introduction, it shows the nonperturbative behavior $\Delta\propto \exp(-2/g\rho)$ which cannot be obtained by the perturbation in $g$ since the convergence radius of the Maclaurin expansion of $\exp(-1/x)$ around $x=0$ is zero. 
The energy cutoff introduced here is corresponding to the band width or Debye energy, 
where the former is the natural energy cutoff because of the lattice structure and the latter is the energy cutoff below which the attractive interaction is absent. 
The case of cold atoms is bit different; the s-wave scattering length is used for the regularization. 
After the regularization, the limit $\Lambda\to\infty$ can be harmlessly taken.  

The model in this section is widely applicable in various systems since the zeroth approximation of any interaction is hardcore interaction if the wavelength is longer enough compared with the range of the interaction. 
Moreover, the attractive interaction is renormalized to be stronger in the low temperature region while the repulsive interaction is not, thus the small attractive interaction is sufficient to induce the superconductivity in the temperatures much lower than the binding energy $\Delta$. 
Here, we assumed that the Cooper pairs are formed by the electron pairs whose total momentum and spin are zero. 
Within these assumptions, the pair potential is homogeneous and scalar (which becomes multicomponent for the case of nonzero spin pairing). 
In the above argument, we have used the mean field approach. 
Beyond the mean field approximation, for instance, the model is exactly solved by the method in the integrable system \cite{Richardson1,Richardson2,Gaudin,Cambiaggio}, 
the conformal structure is also investigated \cite{Sierra}. 
In this review, we employ the U(1) breaking scenario to make the comparison with the NJL model clear. 
Alternatively one can consider the number conserving superconducting state where the ground state is given by the U(1) symmetric wave function 
in which the phase of the order parameter is integrated out \cite{LeggettBook}. 
As a result, the gap equation and the energy spectrum are the same with those obtained in this section.

\subsection{Discrete model}
In this subsection, we consider the discretization of the model introduced in the previous subsection. 
The discretized model can be seen from various point of view. 
It can be seen as the discretized approximation naturally required by the numerical simulation. 
One can also consider it as a model of real materials in which an electron moves on the crystal structure formed by the background atoms. 
Alternatively, one can identify it as a model of the cold atom system with an optical lattice, where the lattice structure is formed by the external field made by the optical technique. 

\begin{figure}
\includegraphics[width=7 cm]{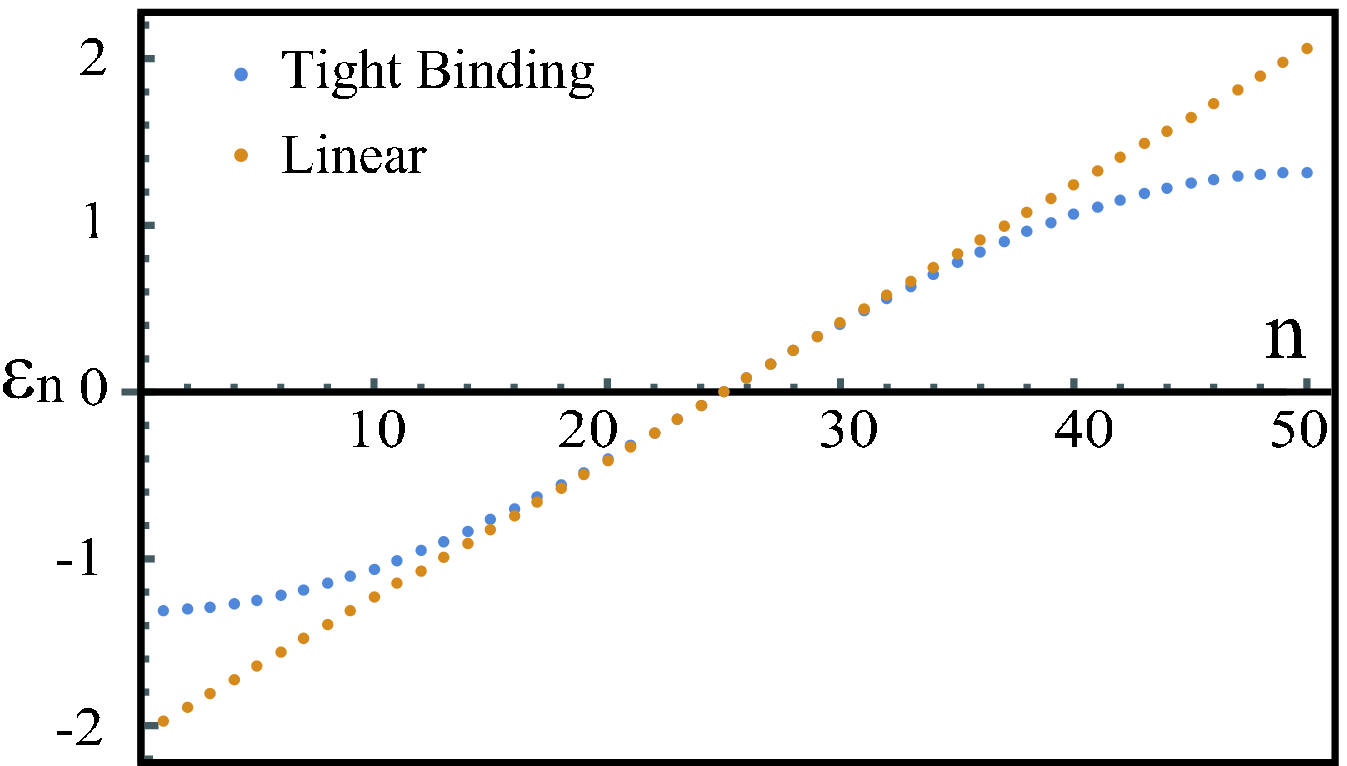}
\includegraphics[width=7 cm]{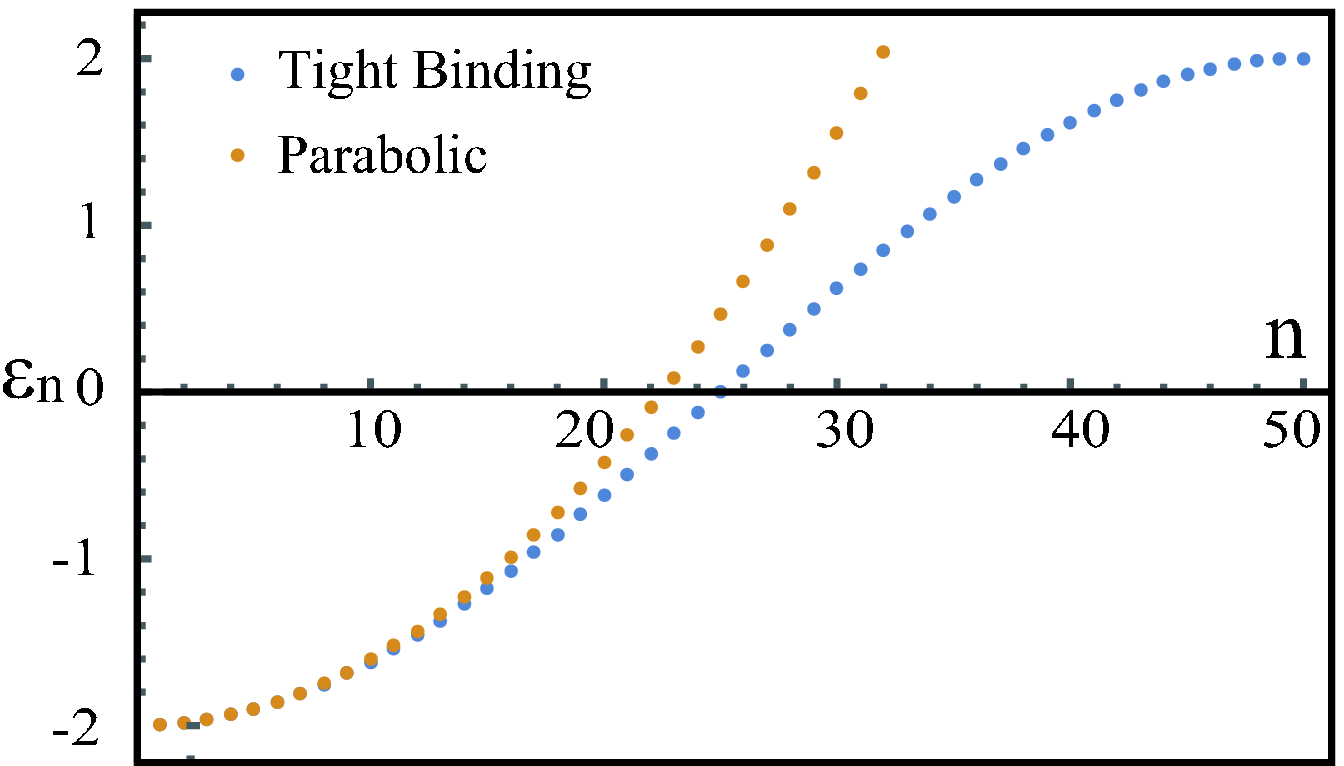}
\caption{The energy spectrum for the tight binding model with $100$ sites. 
The half region of the momentum space $0<k<\pi/a$ is shown. 
The momentum and the number $n$ are related via $k_n=\pi n/L$ with $L=Na$.
The half filling case (left figure) approximates the linear dispersion 
while the small filling case (right figure) approximates the parabolic dispersion. 
If the characteristic energy, such as the superconducting gap, is small enough compared with the band width, the discretized model approximates the continuous system well.}
\label{figtbs}
\end{figure}

First, we discretize the kinetic energy part. 
The differentiation can be replaced by the finite difference. 
Thus, we obtain 
\begin{equation}
H_0=-t \sum_{i,\sigma}\left(c^\dagger_{i+1,\sigma} c_{i,\sigma}+c^\dagger_{i,\sigma} c_{i+1,\sigma}\right),
\label{HKin}
\end{equation}  
where $t$ stands for the transfer integral between the nearest neighbor. 
This Hamiltonian is called the tight binding model since this is the model Hamiltonian for the electrons tightly bind bound at atomic sites in a crystal. 
The transfer integral describes the quantum tunneling process from site to site. 
By the Fourier transformation 
\begin{equation}
c_{n,\sigma}=\frac{1}{\sqrt{N}} \int dk e^{-ikna} c_{k,\sigma},
\label{FourierTrans}
\end{equation}
the first term of the Hamiltonian, which corresponds to a noninteracting Hamiltonian of the electrons, becomes 
\begin{equation}
H_0=
\sum_{k,\sigma} (-2t\cos ka)c_{k,\sigma}^\dagger c_{k,\sigma},
\end{equation}
where $a$ is a lattice constant and $N$ is the number of the total sites. 
Thus, the cosine band structure $\epsilon_k=-2t\cos ka$ is obtained.   
In this model, the band width becomes $4t$. 
This model approximates both the nonrelativistic and relativistic cases well in the continuum limit with large $t$, depending on the filling. 
As shown in Fig.~\ref{figtbs}, the energy dispersion is similar to the linear (relativistic) dispersion in the case of half filling; $E_{\mathrm{F}}=0$ (left figure), 
while the case of small filling; $E_{\mathrm{F}}\sim -2t$, the energy dispersion mimics the parabolic (nonrelativistic) dispersion (right figure).
The hardcore interaction corresponds to the onsite interaction and the chemical potential part corresponds to the energy sift. We thus obtain 
\begin{equation}
H=
-t \sum_{i,\sigma}\left(c^\dagger_{i+1,\sigma} c_{i,\sigma}+c^\dagger_{i,\sigma} c_{i+1,\sigma}\right)
-\mu \sum_{i,\sigma}c^\dagger_{i,\sigma} c_{i,\sigma}
-g\sum_i n_{i,\uparrow}n_{i,\downarrow},
\end{equation}
where $n_{i,\sigma}=c^\dagger_{i,\sigma}c_{i,\sigma}$ is the particle number operator. 

The discretized version of the mean field approximated Hamiltonian is given by 
\begin{equation}
H=
-t \sum_{i,\sigma}\left(c^\dagger_{i+1,\sigma} c_{i,\sigma}+c^\dagger_{i,\sigma} c_{i+1,\sigma}\right)
-\mu \sum_{i,\sigma}c^\dagger_{i,\sigma} c_{i,\sigma}
+\sum_i\Delta_i c^\dagger_{i,\uparrow}c^\dagger_{i,\downarrow}+\sum_i\Delta_i^\ast c_{i,\downarrow}c_{i,\uparrow},
\end{equation}
with the gap equation 
\begin{equation}
\Delta_i=-g\langle c_{i,\downarrow}c_{i,\uparrow} \rangle.
\end{equation}
One needs to solve those equations in a self-consistent manner. 

\subsection{Conducting polymers}
\begin{figure}
\includegraphics[width=12cm]{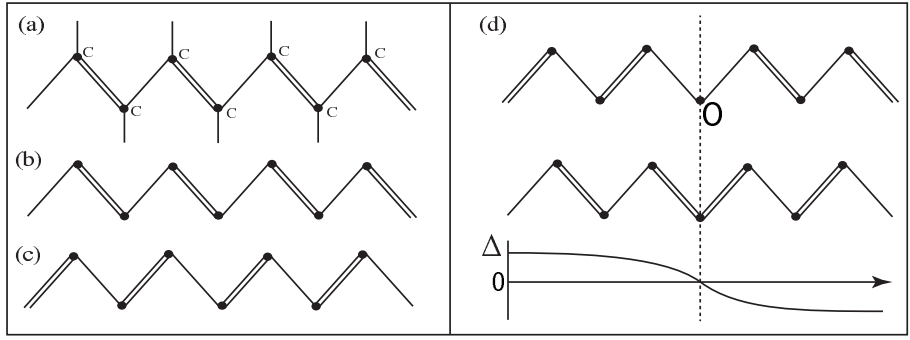}
\caption{The schematic structure of the polyacetylene. 
The black dots stand for the carbon sites. 
For simplicity we omit the hydrogen atoms attached to the end of the vertical lines (Fig.~a). 
The single and double bonds consist of the $\sigma$ electrons and $\sigma+\pi$ electrons, respectively. 
There are right-double pattern (Fig.~b) and left-double pattern (Fig.~c) whose energy are completely degenerate. 
The crucial difference of left-double and right-double patterns occur when the defect is present (Fig.~d). 
In such a case, the order parameter changes its sign at the defect site.}
\label{polyacetylene}
\end{figure}

In this subsection, we consider a trans-polyacetylene system which has a structure depicted in Fig.~\ref{polyacetylene}. 
The polyacetylene has a one dimensional structure consisting of a long carbon atomic chain. 
The bonds between the neighboring carbons are either single or double. 
In the absence of the ionization, the chain has single-double alternating bonds. 
If there is an excess electron in the chain (ionization), the situation is changed. 
As is shown in Fig.\ \ref{polyacetylene}, the single-double region and double-single region are connected at site O. 
The loss of the electron also results in the point defect at site O. 

The model for this system is given as follows \cite{SSH1}: 
\begin{equation}
H=
-\sum_{i,\sigma} [t-\alpha(u_{i+1}-u_i)](c_{i+1,\sigma}^\dagger c_{i,\sigma}+c_{i,\sigma}^\dagger c_{i+1,\sigma})
+\frac{m}{2}\sum_{i}\dot u_i^2+\frac{m\omega^2}{2}\sum_{i}(u_{i+1}-u_i)^2,
\end{equation}
where $u_n$ is the position of the $n$th carbon site and $m$ and $\omega$ are the mass of the carbon atom and the frequency of the vibration, respectively. 
The operator $c^\dagger, c$ are creation and destruction operator of the electrons, respectively, as before. 
The first term is the hopping term and the distance dependence of the hopping probability is taken into account as the deviation of the transfer integral. 
This first term yields the coupling between the electrons and phonons. 
The latter part is the usual harmonic oscillators and describes lattice vibrations. 
We note that this model does not contains the electron-electron interaction in the Hamiltonian level. 
In this model, the electron-electron interaction is indirectly taken into account through the electron-phonon interaction included in the first term of the Hamiltonian. 

In the following, we focus on the static situation and thus drop the second term from the Hamiltonian. 
Physically, this assumption is justified by the huge time scale difference between carbon atoms and electrons; 
the former is $\sim 10^4$ times heavier than the latter. 
The polyacetylene has one conducting electron at each site, which corresponds to the half filling case, where $k_{\mathrm{F}}a=\pi/2$. 
Since the single and double bonds are expected to be aligned in order, we define the field $\tilde u$ as follows 
\begin{equation}
\tilde u=(-1)^i u_i. 
\end{equation}
Then the Hamiltonian becomes
\begin{equation}
H=
-\sum_{i,\sigma} t(c_{i+1,\sigma}^\dagger c_{i,\sigma}+c_{i,\sigma}^\dagger c_{i+1,\sigma})
-2\sum_{i,\sigma} \alpha (-1)^i \tilde u (c_{i+1,\sigma}^\dagger c_{i,\sigma}+c_{i,\sigma}^\dagger c_{i+1,\sigma})
+2m\omega^2\sum_{i}\tilde u^2.
\end{equation}
The first term of the Hamiltonian which corresponds to a noninteracting Hamiltonian of the electrons is the same with $H_{Kin}$ in Eq.\ (\ref{HKin}) in the previous subsection. 
Thus the cosine band $\epsilon_k=-2t\cos ka$ appears by diagonalizing the first term. 
By substituting Eq.\ (\ref{FourierTrans}) into the second part of the Hamiltonian, we obtain 
\begin{equation}
-\frac{2}{N}\sum_{n,\sigma}\sum_{k,k^\prime} \alpha \tilde u e^{\pm2ik_{\mathrm{F}}na}  [e^{ik(n+1)a}e^{-ik^\prime na} +e^{ikna}e^{-ik^\prime (n+1)a}]
 c_{k,\sigma}^\dagger c_{k^\prime,\sigma},
\end{equation}
where we have used a fact that the Fermi wave number is now set to be $k_{\mathrm{F}}=\pi/2a$ and thus $\exp(\pm 2ik_{\mathrm{F}}na)=(-1)^n$. 
We note that a relation
\begin{align}
&\frac{1}{N}\sum_n e^{\pm 2ik_{\mathrm{F}}na}  [e^{ik(n+1)a}e^{-ik^\prime na} +e^{ikna}e^{-ik^\prime (n+1)a}] \nonumber \\
&=\delta_{k-2k_{\mathrm{F}},k^\prime}e^{ika}+\delta_{k+2k_{\mathrm{F}},k^\prime}e^{ika}
+\delta_{k-2k_{\mathrm{F}},k^\prime}e^{-ik^\prime a}+\delta_{k+2k_{\mathrm{F}},k^\prime}e^{-ik^\prime a},
\end{align}
holds, which means that the second term of the Hamiltonian describes the $\pm 2k_{\mathrm{F}}$ momentum transfer process. 
This process is called the umklapp process. 
The second term of the Hamiltonian $H_{umk}$ becomes 
\begin{align}
H_{umk}=-2\alpha i \sum_{k,\sigma} \tilde u 
[c_{k,\sigma}^\dagger c_{k-2k_{\mathrm{F}},\sigma}e^{ika}
+c_{k,\sigma}^\dagger c_{k+2k_{\mathrm{F}},\sigma}e^{ika}
-c_{k,\sigma}^\dagger c_{k-2k_{\mathrm{F}},\sigma}e^{-ika}
-c_{k,\sigma}^\dagger c_{k+2k_{\mathrm{F}},\sigma}e^{-ika}],
\end{align}
where we again used that $k_{\mathrm{F}}=\pi/2a$ and $\exp(\pm 2ik_{\mathrm{F}}a)=-1$. 
As in the case of the BCS theory, by choosing the zero point of the momentum to be $k_{\mathrm{F}}$, we obtain $\exp(\pm ika)=\exp[i(\pm \tilde k+k_{\mathrm{F}})a]=i \exp(\pm i\tilde k a)$. By rewriting $\tilde k$ to be $k$, we obtain 
\begin{align}
H_{umk}=4\alpha \sum_{k,\sigma} \tilde u\sin ka 
(c_{k,\sigma}^\dagger c_{k-2k_{\mathrm{F}},\sigma}+c_{k,\sigma}^\dagger c_{k+2k_{\mathrm{F}},\sigma}).
\end{align}

It is convenient to introduce the valence band $|k|\le \pi/2a$ and the conduction band $\pi/2a<|k|\le \pi/a$.  
According to this devision, we define 
\begin{equation}
c_{k,\sigma}^v=c_{k,\sigma},\, c_{k,\sigma}^c=c_{k\pm2k_{\mathrm{F}},\sigma},
\end{equation}
for the valence band electrons and the conduction band electrons, respectively. 
Here the $\pm$ of the conducting electrons is chosen such that $k$ should be in the region $\pi/2a<|k|\le \pi/a$. 
The Hamiltonian is now reduced to be
\begin{equation}
H=
\sum_{k,\sigma} |\epsilon_k|(c_{k,\sigma}^{c\dagger} c^c_{k,\sigma}-c_{k,\sigma}^{v\dagger} c^v_{k,\sigma})
+4\alpha \sum_{k,\sigma} \tilde u \sin ka(c_{k,\sigma}^{c\dagger} c^v_{k,\sigma}+c_{k,\sigma}^{v\dagger} c^c_{k,\sigma})
+2m\omega^2\sum_{i}\tilde u^2.
\end{equation}
The electron part of the Hamiltonian has the same structure with the BCS Hamiltonian (\ref{NambuRep}) as follows 
\begin{equation}
H=
\sum_{k,\sigma}\left(
c_{k,\sigma}^{c\dagger} \ c_{k,\sigma}^{v\dagger} 
\right)
\left(
\begin{array}{cc}
|\epsilon_k| & 4\alpha\tilde u \sin ka\\
4\alpha\tilde u \sin ka & -|\epsilon_k|
\end{array}
\right)
\left(
\begin{array}{cc}
c_{k,\sigma}^c \\
c_{k,\sigma}^v 
\end{array}
\right).
\label{SSH}
\end{equation}

By taking the continuum limit and linear dispersion approximation, the above Hamiltonian becomes \cite{TLM}
\begin{equation}
H=
\sum_\sigma \int dx\left(
c_\sigma^{c\dagger}(x) \ c_\sigma^{v\dagger}(x) 
\right)
\left(
\begin{array}{cc}
-i\partial_x & \Delta_{SSH} \\
\Delta_{SSH} & i\partial_x
\end{array}
\right)
\left(
\begin{array}{cc}
c_\sigma^c(x) \\
c_\sigma^v(x) 
\end{array}
\right)
\label{SSHreal}
\end{equation}
in real space. 
Here we have defined $\Delta_{SSH}$ as the inverse Fourier transformation of $4\alpha\tilde u\sin ka$. 
The stationary condition for $\Delta_{SSH}$ gives the following gap equation. 
\begin{equation}
\Delta_{SSH}
=-4\frac{a\alpha^2}{m\omega^2}
\sum_{\sigma} 
\left\langle c_{\sigma}^{c\dagger}(x) c^v_{\sigma}(x)+c_{\sigma}^{v\dagger}(x) c^c_{\sigma}(x)\right\rangle.
\label{SSHgap}
\end{equation}
These equations are the same with the BdG equation and the gap equation obtained from the NJL model. 
In this model, the electron-phonon interaction results in the dimerized order for the lattice in which the size of a unit cell is expanded to be $2a$ from $a$. 
At the same time, the electrons also have the density modulation (charge density wave state) which is equidistant with the distance $2a$. 
 
It should be noted that the case of the BCS theory, the particle and hole with different spin form the two component spinor and the equation for the spinor is the same with one for the NJL model, while in the case of the polyacetylene, the spinor consists of the conduction and valence bands with the same spin. 
It should be also noted that the condensation $\Delta_{SSH}$ for the polyacetylene naturally arises from the vibration of the lattice 
and thus it is real valued while BCS pair potential can be complex valued.

\section{Inhomogeneous solutions}{\label{InhomogeneousSols}}
In the previous section, we have focused on the homogeneous condensates. 
Here we consider inhomogeneous condensates.
\subsection{Kink solution, kink crystal solution, and zero mode}
The first topological soliton has been found in the polyacetylen system \cite{SSH1, TLM} (for more details, see Ref.\ \cite{Heeger} and references therein).
The single-double alternating bond structure of the polyacetylen can have the left-double or the right-double pattern (Fig.~\ref{polyacetylene}). 
These two patterns are not physically different if there is no defect since it is just a matter from which side this material is seen. 
However in the presence of the defect at which the left-double and right-double patterns are connected, 
the system is distinguished from the no defect state in terms of the topology (Fig.~\ref{polyacetylene}). 
In such a case, the condensation has a kink structure which is topologically stable. 
The important point is that the condensate is real for this system, which forbids to transform the solution to a homogeneous solution by local deformations. 
In other words, the left-double and right-double patterns are distinguished by the $\mathbb{Z}_2$ index. 
As a consequence, the order parameter must be zero at the defect point since the continuous transition from left-double to right-double patterns is forbidden. 
It can be shown that the following kink solution satisfies the BdG equation and the gap equation self-consistently: 
\begin{equation}
\Delta(x)=m \tanh mx. 
\label{kinksol}
\end{equation} 
In order to check the self-consistency, it is convenient to choose the different representation. 
In the previous section, we have chosen the representation $\mathcal{H}=-i\sigma_z \partial_x+\sigma_x \Delta$. 
Instead, here we choose $\mathcal{\tilde  H}=-i\sigma_y \partial_x+\sigma_x \Delta$., where $\vec\sigma$ is the Pauli matrix vector. 
Here and hereafter we denote the Hamiltonian density by $\mathcal{H}$. 
This canonical transformation is achieved by the unitary transformation $U^\dagger \mathcal{\tilde H} U$ with $U=(1-i\sigma_x)/\sqrt{2}$. 
The BdG equation in this representation reads 
\begin{equation}
\mathcal{\tilde H} 
\left(
\begin{array}{c}
\psi_\sigma^u(x) \\
\psi_\sigma^l(x) 
\end{array}
\right)
=
\left(
\begin{array}{cc}
0 & -\partial_x + \Delta(x) \\
\partial_x + \Delta(x) & 0
\end{array}
\right)
\left(
\begin{array}{c}
\psi_\sigma^u(x) \\
\psi_\sigma^l(x) 
\end{array}
\right)
=
E
\left(
\begin{array}{c}
\psi_\sigma^u(x) \\
\psi_\sigma^l(x) 
\end{array}
\right).
\label{BdGChiral}
\end{equation}
This Dirac-like equation can be rewritten as the following diagonal Klein-Goldon(KG)-like equation by calculating $\mathcal{\tilde H}^2$ 
\begin{equation}
E^2
\left(
\begin{array}{c}
\psi_\sigma^u(x) \\
\psi_\sigma^l(x) 
\end{array}
\right)
=
\left(
\begin{array}{cc}
-\partial_x^2 + \Delta(x)^2-[\partial_x \Delta(x)] & 0 \\
0 & -\partial_x^2 + \Delta(x)^2+[\partial_x \Delta(x)]
\end{array}
\right)
\left(
\begin{array}{c}
\psi_\sigma^u(x) \\
\psi_\sigma^l(x) 
\end{array}
\right).
\label{KGeq}
\end{equation}
Substituting the kink solution Eq.\ (\ref{kinksol}) into the above KG equation, one can show that the lower component satisfies 
$(-\partial_x^2+m^2)\psi_\sigma^l(x)=E^2\psi_\sigma^l(x)$, 
which is the same differential equation with the one for the homogeneous condensate. 
Thus, the energy spectrum becomes $E_k=\pm \sqrt{k^2+m^2}$ which is the same with the one for the homogeneous condensation except for the zero energy state. 
The zero energy state is obtained by $(\psi_{0,\sigma}^u, \psi_{0,\sigma}^l)\propto (\exp[-\int^x \Delta(x)dx],0)$ or 
$(\psi_{0,\sigma}^u, \psi_{0,\sigma}^l)\propto (0,\exp[\int^x \Delta(x)dx])$ if and only if the mode is normalizable \cite{Jakiew-Rebbi}. 
In the case of homogeneous condensate, neither of these modes are normalizable, 
while in the case of the kink solution the former is normalizable and yields 
$(\psi_{0,\sigma}^u, \psi_{0,\sigma}^l)= (\sqrt{m/2}\cosh(mx),0)$. 
Since the energy spectrum is completely the same with the one for the homogeneous condensate and the zero energy state does not enter to the gap equation since it does not change the total energy at all, $m$ coincides with the amplitude of the gap for the homogeneous condensate. 
This is physically obvious since at sites infinitely far from the defect site, the condensation should be the same with the one without the defect. 
By the same reason, the energy spectrum with this kink condensate should have almost the same structure with the one without the defect. 
The only difference is that in the case of the kink condensate there is a zero energy mode (mode at the Fermi energy). 

\begin{figure}
\includegraphics[width=12cm]{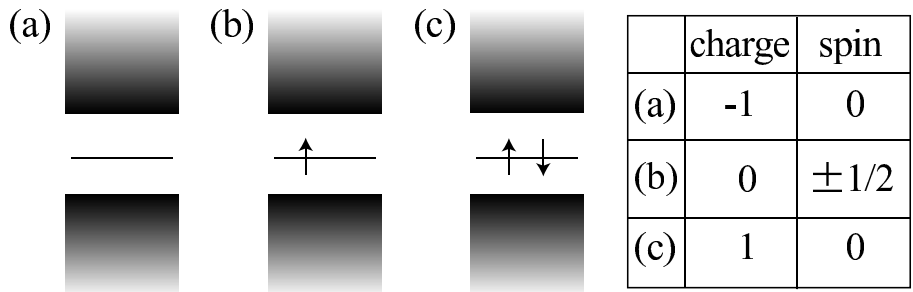}
\caption{The occupation number of the zero mode and the charge-spin separation. 
Fig.~(a), (b), and (c), respectively, show the empty state, the singly occupied state, and the doubly occupied state.
Those situations yield the charge-spin separation of the conducting degrees of freedom.}
\label{SCsep}
\end{figure}

It is known that the charge neutral state in this solution requires the one electron occupation in the zero mode \cite{SSH2}. 
In such a case, the total spin becomes $\pm 1/2$ depending on the spin of the zero mode electron (Fig.~\ref{SCsep} (b)). 
On the other hand, if the zero mode is empty, the total spin is $0$ while the charge is $+1$ in the unit of $|e|$ (Fig.~\ref{SCsep} (a)). 
Similarly, if the zero mode is doubly occupied, the total spin is $0$ while the charge is $-1$ (Fig.~\ref{SCsep} (c)). 
Thus, the interesting feature in this system can be observed from these facts. 
In this system, the charge $0$, spin $\pm 1/2$ transport or the charge $\pm 1$, spin $0$ transport can be realized 
when the kink contributes to the transport phenomena. 
Indeed, the nonzero spin for the charge neutral soliton in the conducting polymer is observed experimentally \cite{Weinberger, Nechtschein} (for more detail, see Ref.~\cite{ConductingPolymar}). 
In the case of larger internal degrees of freedom, the fractional charges is also predicted \cite{Jakiew-Rebbi,GoldstoneWilczek,Niemi-Semenov}. 

In the above discussion, we have focused on the single defect for simplicity. 
We move to the multiple defects case.  
At each defect, the left-double and right-double patterns are changed and the zero mode wave function has a configuration having a peak at each site. 
If the zero mode is completely occupied, each defect has a negative charge, while if it is completely empty, 
each defect has a positive charge and thus the Coulomb repulsion is expected to stabilize the state with equidistant kink-antikink configuration. 
The analysis of the dynamical stability requires fully time dependent treatment and that is not discussed here. 
Instead, we discuss about the static self-consistent solutions. 
The kink-anti kink crystalline solution is given as \cite{Brasovskii,Horovitz}
\begin{equation}
\Delta(x)=m \mathrm{sn} (mx, \nu), 
\label{kinkcrystalsol}
\end{equation} 
where $\mathrm{sn}$ is the Jacobi's elliptic function and $\nu$ is the elliptic parameter. 
In this state, a unit cell consists of a kink-anti kink-kink and is larger than that of the uniform dimerized state where the size of the unit cell is $2a$. 
As a result, the zero energy mode appearing in a single kink solution becomes the inner gap continuum whose width is determined by the density of the kink-anti kink lattice characterized by the elliptic parameter $\nu$ \cite{Brasovskii}. 
The single kink limit $\nu\rightarrow 1$ reproduces the zero mode isolated in the gap. 

Since the theory of the superconductor shares the same self-consistent equations, the BdG equation and the gap equation, with the polyacetylene system, 
the above inhomogeneous solutions are also solutions in the superconducting phase. 
The inhomogeneous solutions have been proposed in the presence of the spin imbalance for the superconductor. 
In the spin imbalance system, the size of the Fermi surface for the upper spins and lower spins are different and 
thus the Cooper pair have a non-zero momentum which results in the inhomogeneous condensate. 
This scenario has been proposed by Fulde-Ferrell \cite{FF} and Larkin-Ovchinnikov \cite{LO} independently. 
The analysis corresponding to the above kink crystal solution for the superconductors have been done in Ref.~\cite{BarSagi}. 
A similar situation is realized by a cold atom system where the spin is replaced by the quasi spin played by the different inner state of the atom. 
The periodic pattern of spin accumulation is indeed observed in the cold atom system and is considered as the evidence of the kink crystal state \cite{Liao}. 

\subsection{Nonlinear Schr\"odinger equation and various inhomogeneous solutions}
\begin{figure}
\includegraphics[width=12cm]{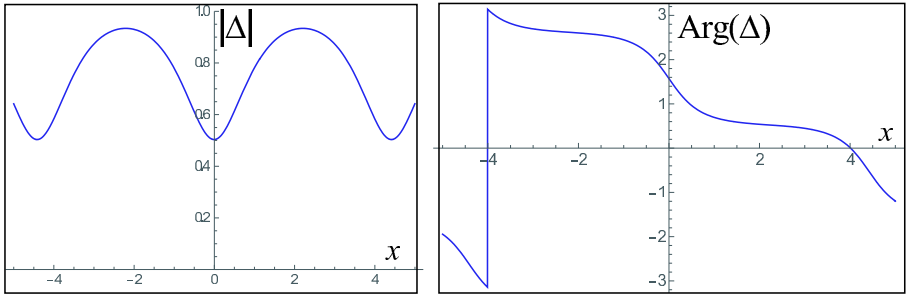}
\caption{The general solution of the NLSE. Both the amplitude and the phase modulate in space. 
The amplitude of the condensate does not touch to zero in the presence of the current. 
The modulation of the phase is also not uniform in the space.}
\label{FFLO}
\end{figure}

In the previous subsection, we have discussed the kink solution, the kink crystal solution and their zero modes. 
Those solutions are shown to be self-consistent solution to the BdG and the gap equation. 
For those solutions, by a physical intuition a $\tanh$ shape solution and a periodic extension of that solution is found. 
However, more general solutions satisfying both the BdG and gap equations are difficult to find in general. 
More recently, a novel technique was proposed in which the nonlinear Schr\"odinger equation (NLSE) for $\Delta$ is found \cite{Basar-Dunne}. 
The main benefit of this technique is that the NLSE only contains $\Delta$ and few parameters, and does not directly require the fermionic (spinor) solutions. 
Thus, the self-consistent calculation can be avoided. 
The NLSE is given as 
\begin{equation}
\partial_x^2\Delta(x)+i(b-2E)\partial_x \Delta-2(a-Eb)\Delta-2\Delta|\Delta|^2=0,
\label{NLSE}
\end{equation} 
where $a$ and $b$ are parameters which do not depend on the space-time coordinates. 
In the case of a real condensate, 
one can easily show that $b=2E$ and thus the deviation from $b=2E$ roughly corresponds to a phase modulation. 
Here, we just briefly trace the main idea for the derivation for this equation. 
The 1+1 dimensional Dirac system is representable by using the $2\times 2$ Pauli matrices. 
The corresponding spinor consists of two components and two linear independent solution exist. 
Let us call them $\psi_1$ and $\psi_2$ and denote their upper and lower components by $\psi_i^u$ and $\psi_i^l$, respectively. 
The most important observation is that the resolvent $R\equiv \langle x|(H-E)^{-1}|x\rangle$ is describable by those two linear independent solutions as 
\begin{equation}
R(x,E)=\frac{1}{2iW}\left(\psi_1\psi_2^T+\psi_2\psi_1^T\right)\sigma_1, 
\label{resolvent}
\end{equation}
where $W=i\psi_1^T\sigma_2\psi_2$ is the Wronskian of the two independent solutions. 
The resolvent satisfies the Eilenberger equation \cite{Eilenberger}
\begin{equation}
\partial_x R\sigma_3=i\left[
\left(
\begin{array}{cc}
E&-\Delta\\
\Delta^\ast&-E
\end{array}
\right)
,R\sigma_3\right]. 
\label{DEeq}
\end{equation}
This equation can be derived from the BdG equation and Eq.~(\ref{resolvent}). 
The resolvent also satisfies the Hermitian condition $R^\dagger=R$, the traceless condition $\mathrm{Tr}R\sigma_3=0$, 
and the constant determinant condition $\det R=-1/4$. 
Together with the gap equation, one can impose an ansatz for the resolvent as 
\begin{equation}
R=\left(
\begin{array}{cc}
a+|\Delta|^2&b\Delta-i\partial_x\Delta\\
b\Delta^\ast +i\partial_x\Delta^\ast&a+|\Delta|^2
\end{array}
\right).
\end{equation}
This ansatz is, by construction, consistent with the gap equation, and the substitution of this ansatz into the Eilenberger equation (\ref{DEeq}) yields the NLSE (\ref{NLSE}).
The most general solution for the above equation is given by 
\begin{equation}
\Delta=-A\frac{\sigma(Ax+iK^\prime-i\theta/2)}{\sigma(Ax+iK^\prime)\sigma(i\theta/2)}
\exp\left\{iAx[-i\zeta(i\theta/2)+i\mathrm{sn}(i\theta/2)^{-1}]+i\theta\eta_3/2\right\},
\end{equation}
where $\sigma,\ \zeta$ are the Weierstrass sigma and zeta functions, respectively. 
The half-periods $\omega_1$ and $\omega_3$ for real and imaginary directions are set to be $\omega_1={K}$ and $\omega_3=i{K}^\prime$, 
with the complete elliptic integral of the first kind ${K}(\nu)=\int_0^{\pi/2}dt(1-\nu\sin^2 t )^{-1/2}$ and ${K}^\prime\equiv {K}(1-\nu)$. 
The constant $\eta_3$ is defined by $\eta_3=\zeta (i{K}^\prime)$. 
The scale of the condensate is set by the the parameter $A$ as $A=-2im\mathrm{sc}(i\theta/4)\mathrm{nd}(i\theta/4)$. 
Here $\mathrm{sc}=\mathrm{sn}/\mathrm{cn}$ and $m, \theta$ are related to the amplitude and the phase modulation, respectively. 
In Fig.\ \ref{FFLO}, we plot the general solution which shows the modulation of the amplitude and phase. 

Here, we mention few points before closing this section. 
This method is also generalized to the time-dependent case \cite{Klotzek,Fitzner,Basar-Dunne2,Dunne,Dunne2, Efimkin}. 
There are other theoretical schemes to isolate $\Delta$ from the fermionic degrees of freedom to avoid the self-consistent calculation. 
One of them is the Ginzburg-Landau (GL) expansion which contains few phenomenological parameters to be exported from elsewhere. 
This method is valid in the vicinity of the critical point since this method is basically the small $\Delta$ expansion. 
The dimensional analysis also allows one to include the derivative terms and the inhomogeneous solution is obtained \cite{Buzdin}. 
The method of the effective Lagrangian is also suitable to obtain a equation containing only $\Delta$ \cite{Kunihiro-Hatsuda}. 
In the case of the GL expansion and the effective Lagrangian method, 
the equation for $\Delta$ is directly derived from the microscopic theory with the physically reasonable ansatz. 
It should be noted that the method explained in this section is applicable in any temperature even far from the critical point and 
the gradient expansion is not used.

\subsection{Internal structures behind NJL system}\label{INTSUSY}
In the previous section, we have briefly shown the method to avoid self-consistent calculations. 
Here we show the extension of the method by using the integrable structure behind \cite{CDP, TTYN, TakahashiNitta}. 
By rewriting the Eilenberger equation (\ref{DEeq}) as 
\begin{equation}
\partial_t U-\partial_x V+[U,V]=0,\quad  \partial_t V=0,
\end{equation}
with 
\begin{equation}
U=\left(
\begin{array}{cc}
iE&-i\Delta\\
i\Delta^\ast&-iE
\end{array}
\right),\quad  
V=R\sigma_z
\end{equation}
one can see this equation is nothing but the integrable condition (zero curvature condition). 
The spinor obeys the following equation
\begin{equation}
\partial_x \psi=U\psi. 
\end{equation}
Those equations are solvable by using the integrable structure, for instance, using the inverse scattering method. 
The present problem is categorized to the Ablowitz-Kaup-Newell-Segur (AKNS) hierarchy \cite{AKNS}. 
The method in the previous subsection corresponds to the AKNS$_1$ 
while the minimal model to describe the constant solution is belonging to AKNS$_{-1}$. 
By elevating to the higher hierarchy, one can obtain more general solutions, e.g., a kink-antikink bound state \cite{DHN, Campbell} and 
kink-anti kink-kink solution \cite{OkunoOnodera, Feinberg}.

There is another remarkable structure in the 1+1 dimensional NJL model, namely supersymmetry. 
In the previous subsection, the Klein-Gordon-like equation (\ref{KGeq}) has been derived from the Dirac-like equation (\ref{BdGChiral}). 
The Hamiltonian density operator $\mathcal{\tilde H}$ can be decomposed into $\mathcal{\tilde H}=A^++A^-$ 
\begin{equation}
A^+
=\left(
\begin{array}{cc}
0&Q^+\\
0&0
\end{array}
\right),\quad  
A^-
=\left(
\begin{array}{cc}
0&0\\
Q^-&0
\end{array}
\right),\quad  
Q^\pm=\mp \partial_x+\Delta.
\end{equation}
This represents a supersymmetric structure. 
The operators $A^\pm$ and the gap $\Delta$  can be interpreted as a supercharge and superpotential, respectively \cite{FeinbergS, Cooper-Khare-Sukhatme}. 
The operator appearing in the Klein-Gordon-like equation is easily obtained 
by $\mathcal{\tilde H}^2=(A^++A^-)^2=A^+A^-+A^-A^+=\mathrm{diag}(Q^+Q^-, Q^-Q^+)$. 
Since $[\mathcal{\tilde H}^2, Q^\pm]=0$, the upper and lower components of the Klein-Gordon-like equation yield the same energy spectrum except for the zero mode. 
We partially utilized this benefit in the calculation of the energy spectrum for the kink solution, 
where the one component of the Klein-Gordon-like equation coincides with the one for the constant condensate. 
In this case, we independently calculate the zero mode by using the general solution for the zero mode with the normalizable condition. 

The zero mode(s) is (are) given as the solution(s) of $Q^+\psi_0=0$ and/or $Q^+\psi_0=0$: 
\begin{equation}
\psi_0=A\exp \left(\pm \int^x dx\Delta(x) \right),
\end{equation}
where $A$ is the integral constant. 
If and only if the above solution(s) is (are) normalizable, the zero mode(s) can exist. 
This zero mode yields the superpotential as 
\begin{equation}
\Delta=\pm \frac{\partial_x \psi_0}{\psi_0}.  
\label{ZeromodeSP}
\end{equation}
This structure is useful to find new solutions. 
For instance, the superpartner of the constant potential system with the Dirichlet boundary condition can be used to find a nontrivial condensate 
with which all fermionic solutions satisfy the Dirichlet boundary condition \cite{FNTY}.

\subsection{Inhomogeneous solutions on a ring}
\begin{figure}
\includegraphics[width=4.5cm]{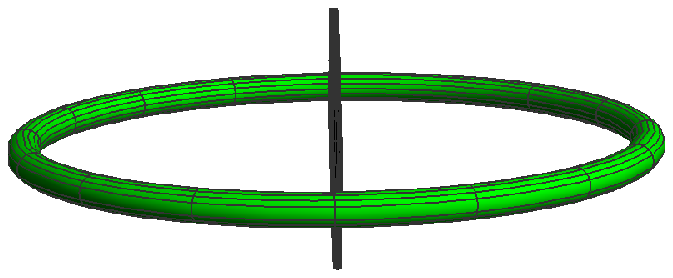}
\includegraphics[width=4.5cm]{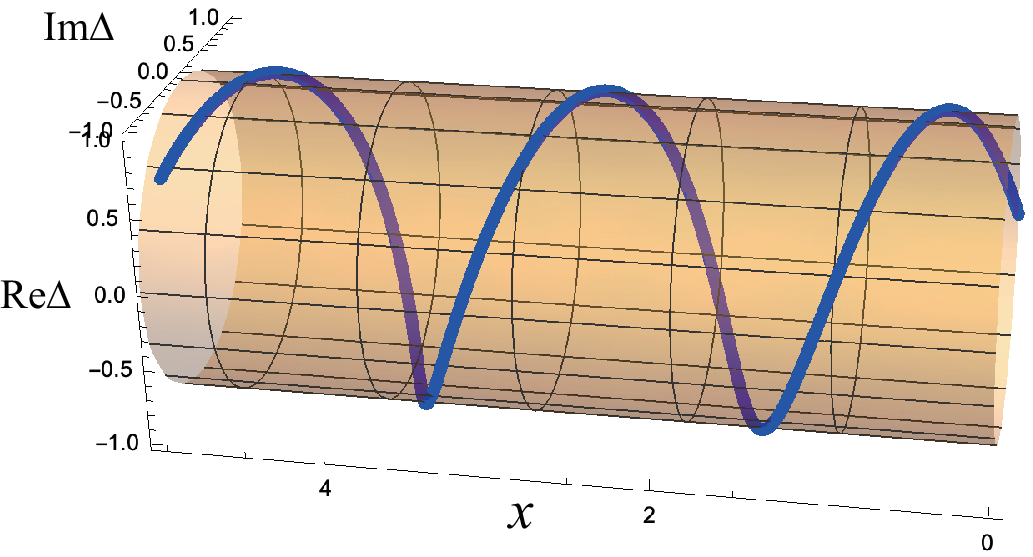}
\includegraphics[width=4.5cm]{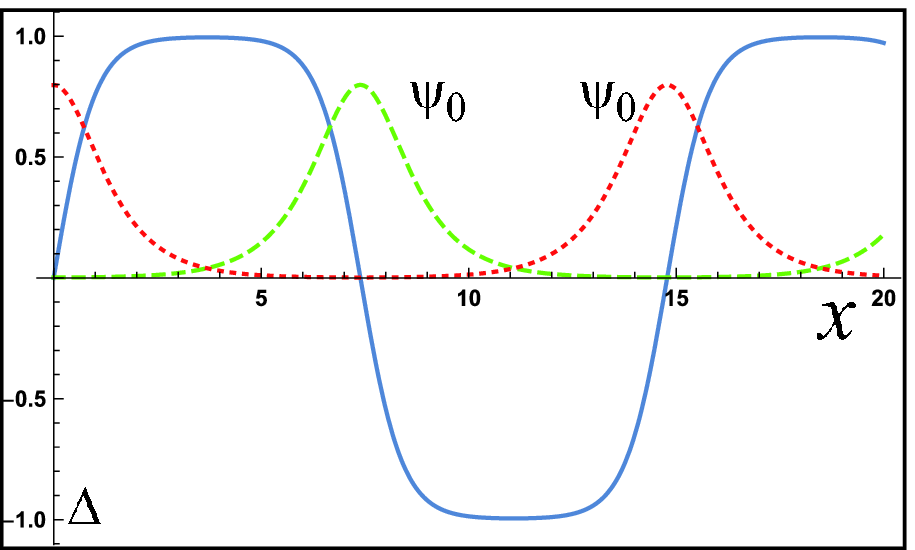}
\caption{The model considered in this subsection. 
The magnetic flux (black line) penetrating the supeconducting ring. 
The Zeeman field is also applied on the ring. 
A typical configuration of the FF phase (middle figure) and that of the LO phase (right figure) with $\nu=0.99$ are shown. 
The horizontal and vertical axes correspond to the spatial coordinate and $\Delta$, respectively. 
The solid line represents the order parameter and the dotted and dashed lines are the configuration of the zero modes, 
which is scaled as it fits to the figure. 
 }
\label{ring}
\end{figure}

In this subsection, we discuss the stability of the inhomogeneous solutions. 
In the case of the polyacetylene, the kink structure appears by putting the defect which corresponds to the injection of the dopant in an experiment. 
Here, we consider a possibility of the dynamical formation of the inhomogeneity for a superconductor/superfluid. 
As we mentioned before, the spin imbalance induces the nonzero total momentum to Cooper pairs \cite{Machida-Nakanishi,footnote4}.
This imbalance can be achieved by applying a magnetic field on superconductors. 
The same situation is realizable in cold atom systems by tuning the population of the spin-up and spin-down particles to be different. 
The phase modulation of the superconducting state brings the current flow to superconductors. 
This slope of the phase can be induced by using the Aharonov-Bohm (AB) effect. 
Thus, we consider the setup depicted in Fig.\ \ref{ring}. 
The superconducting ring is penetrated by the magnetic flux $\Phi$ which can tune the phase winding. 
In addition, we also apply the magnetic field on the superconductor, which induces the spin imbalance. 
In order to avoid the confusion, we refer the latter magnetic field as the Zeeman field. 

Unlike the analytic solutions in the previous subsections, we introduce self-consistent numerical solutions as follows. 
We employ the above discretized model with the external magnetic fields 
\begin{align}
H=&
-\sum_{i,\sigma}\left(t e^{i\phi a/L}c^\dagger_{i+1,\sigma} c_{i,\sigma}+t e^{-i\phi a/L}c^\dagger_{i,\sigma} c_{i+1,\sigma}\right)\nonumber\\
&+\sum_{i,\sigma}\sigma h c^\dagger_{i,\sigma} c_{i,\sigma}
-\mu \sum_{i,\sigma}c^\dagger_{i,\sigma} c_{i,\sigma}
+\sum_i\Delta_i c^\dagger_{i,\uparrow}c^\dagger_{i,\downarrow}+\sum_i\Delta_i^\ast c_{i,\downarrow}c_{i,\uparrow},
\end{align}
where $\phi=2\pi \Phi/\Phi_0$ with the magnetic flux quanta $\Phi_0=hc/e$, is so-called the Peierls phase especially for the discretized model or AB phase in general. 
The first term contains the effect of the flux, where the hopping has the phase depending on the magnetic flux penetrating the ring. 
As the well known AB effect, if and only if the magnetic flux is enclosed by the system, 
the physical quantity such as a energy spectrum depends on the AB phase. 
The second term together with the third term represent the sift of the chemical potential by the Zeeman field. 
The BdG and gap equations derived from this model become 
\begin{equation}
\left(
\begin{array}{cc}
H_{i,,j,\uparrow}&\Delta_i \delta_{i,j}\\
\Delta^\ast_i\delta_{i,j}&-H_{i,j,\downarrow}^\ast
\end{array}
\right)
\left(
\begin{array}{c}
u_{j,\uparrow}\\
v_{j,\downarrow}
\end{array}
\right)
=E\left(
\begin{array}{c}
u_{i,\uparrow}\\
v_{i,\downarrow}
\end{array}
\right),
\end{equation}
and
\begin{equation}
\Delta_i=g^2\sum_E u_{i,\uparrow}(E) v_{i,\downarrow}^\ast(E) \tanh \frac{E}{2T},
\end{equation}
where $H_{i,j,\sigma}=-te^{2\pi i\phi} \delta_{i,i+1}-te^{-2\pi i\phi} \delta_{i,i-1}+(\sigma h-\mu) \delta_{i,j}$ and $(u_n v_n)$ is the $n$-th eigen spinor of the Hamiltonian. 
Here, we have introduce the temperature $T$ which is ignored in the previous sections. 
One needs to solve these equations self-consistently. 
In the following calculation, we prepare various initial conditions 
such as $\Delta_n\propto e^{ikna}$ or $\Delta_n \propto \sin kna$, or $\Delta_n \propto e^{ikna} \sin k^\prime na$ with various $k$, $k^\prime$.  
For the stability, we compare free energies 
\begin{equation}
F=-T\sum_E \ln \left(1+e^{-E/T}\right)+\frac{1}{2g}\sum_i|\Delta_i|^2-\sum_i (\mu+h),
\end{equation}
of the various solutions and seek the state with the smallest free energy. 

\begin{figure}
\includegraphics[width=10cm]{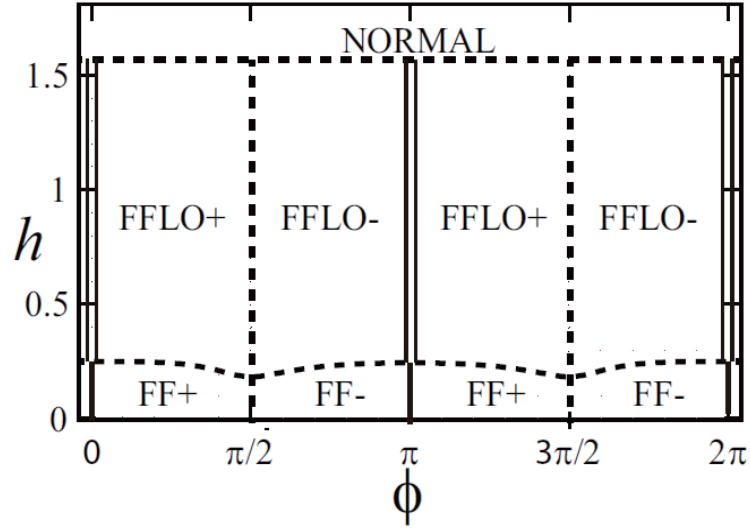}
\caption{The phase diagram of the supercoducting ring. 
The solid and double solid lines correspond to the constant condensation (BCS phase) and the LO phase, respectively. 
The FF phase and FFLO phase are classified to the two classes according to the direction of the current flow ($+$ for clockwise and $-$ for counterclockwise). 
The dotted lines represent first order phase transition lines. 
 }
\label{phasediagram}
\end{figure}

It is shown by numerical calculations that the magnetic flux stabilizes the phase modulating solution (FF state), 
while the magnetic field on the ring results in the condensate with the amplitude modulation (LO state) \cite{Quan, Yoshida1}. 
This result can be understood as follows. 
The magnetic field on the superconductor increases the excess spins which cannot form the Cooper pair. 
These excess spins increase the total energy in the presence of the condensate while the condensate of Cooper pairs lower the total energy. 
As a consequence of this competition, the partially broken condensate is favored between the normal state and the constant condensate by changing the strength of the Zeeman magnetic field  \cite{Machida-Nakanishi}. 
On the other hand, the magnetic flux does not induce the spin imbalance though it induces the total momentum to the Cooper pairs. 
Thus, the magnetic flux results in the condensate with a finite current without the amplitude modulation. 
Numerically, it is found that the interplay of the magnetic flux and Zeeman magnetic field results in the condensate with both phase and amplitude modulations (FFLO state) as shown in Fig.~\ref{phasediagram} \cite{YTTMHN}. 
It is also shown that the Basar-Dunne's method can be generalized in the presence of the magnetic fields \cite{YTMN}. 
Thus, FFLO states found in the numerical calculation can be identified as discrete versions of the twisted kink crystal state. 

The other possibility to dynamically induce the inhomogeneous solution is a bilayer system with the spin-orbit coupling. 
In such a case, the spatially modulated condensation is found to be stable \cite{Yoshida-Sigrist-Yanase}. 
The finite density also drives the system into the inhomogeneous state \cite{Nickel}. 
The generalization to the multi-band system is also considered \cite{Mizushima}. 

\subsection{Finite System and Casimir Force}
\begin{figure}
\includegraphics[width=12cm]{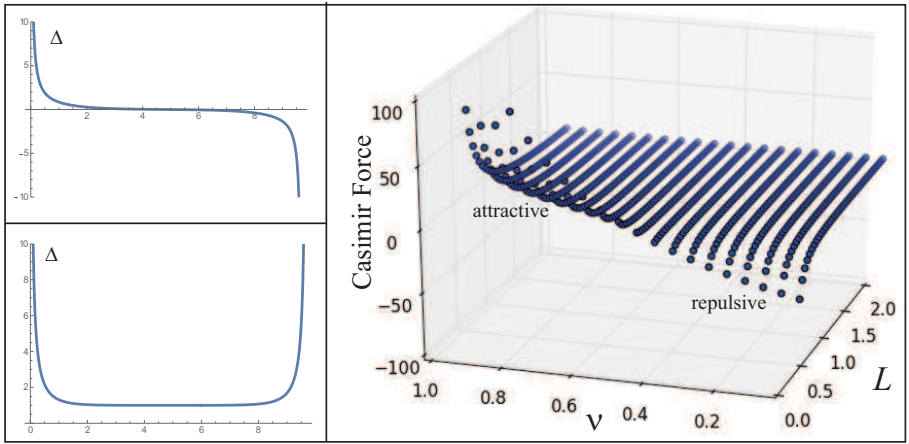}
\caption{The normal-like solution (top left) and the BCS-like solution (bottom left figure) for $nu=0.999$. 
Both the solutions diverge towards the boundaries. 
The Casimir force for the BCS-like solution changes the sign by changing $\nu$ (right).} 
\label{FiniteNJL}
\end{figure}

The method can also be used to the case of condensates in a finite interval with the Dirichlet boundary condition \cite{FNTY} 
which mimics heavy impurities embedded in a superconductor or cold atom system. 
In this situation, the vacuum fluctuation gives the pressure between the two impurities. 
By using the method of the supersymmetry (Sec.\ \ref{INTSUSY}), one solution can be found as follows. 
The Klein-Gordon equation (length $L$) with the Dirichlet boundary condition in the presence of the constant potential $V_0$ has 
the eigenstates and eigenenergies as 
\begin{equation}
\psi_n=\sqrt{2}{L}\sin\frac{\pi x}{L},\quad  \epsilon_n=(\pi n/L)^2+V_0, 
\end{equation}
respectively. 
Thus, a zero mode appears if we apply the constant potential $V_0=-\pi^2/L^2$. 
Using this zero mode and the formula (\ref{ZeromodeSP}), one can find the superpotential $\Delta$ 
\begin{equation}
\Delta=-\frac{\pi}{L} \cot \frac{\pi x}{L}. 
\label{cot}
\end{equation}
From the asymptotic analysis in the vicinity of the boundary, it can be shown that the condesation $\Delta$ must diverge as $1/x$ and $1/(L-x)$, 
respectively, near $x=0$ and $x=L$. 
The above solution satisfies this condition and it is also straightforward to show that this solution also satisfies the NLSE (\ref{NLSE}). 
From this solution, it is not difficult to find the general solution which satisfies Eq.~(\ref{NLSE}), 
\begin{equation}
\Delta=m \frac{\mathrm{cn} \left(mx,\nu\right)}{\mathrm{sn} \left(mx,\nu\right)}, 
\label{cs}
\end{equation}
where $m=2K(\nu)/L$ and  $K(\nu)$ is the elliptic integral of the first kind. 
In Fig.\ \ref{FiniteNJL}, we plot the solution for $\nu=0.999$. 
The bulk part corresponds to the normal phase whereas the diverging behavior is observed in the vicinity of the boundary. 
The solution (\ref{cot}) corresponds to the case of $\nu\to 0$. 
By taking another limit (translation $x\to x+K(\nu)$ and $\nu\to 1$), the solution (\ref{cs}) becomes zero everywhere. 
Thus, the solution is considered as a cousin of the normal phase. 

The other class of solutions can be found as 
\begin{equation}
\Delta=m \frac{1}{\mathrm{sn} \left(mx,\nu\right)}, 
\label{sn}
\end{equation}
which corresponds to the BCS solution in a certain limit. 
In Fig.\ \ref{FiniteNJL}, we plot the solution for $\nu=0.999$. 
The bulk part corresponds to the BCS phase whereas the diverging behavior is observed in the vicinity of the boundaries. 
In the right figure of Fig.\ \ref{FiniteNJL}, it is shown that the Casimir force calculated by $F_c=-\partial_L E_{\mathrm{tot}}$ changes the sign by changing $\nu$. 
In the infinite system, $\nu$ should be 1 since the solution should be the BCS solution, 
whereas the $\nu=1$ solution in the finite system corresponds to $m\to \infty$. 
Thus, inevitably, $\nu$ should decrease with shrinking the size of the system. 
As a result, the sign change of the Casimir force is found in the presence of the interaction by changing $\nu$ \cite{FNTY}, 
which is absent for the free fermionic system. 
The Casimir force for interacting fermions was also investigated in higher dimensions and other geometries \cite{Inagaki}.

\section{Correspondence between NJL and NL$\sigma$ models}{\label{NJLandNS}}
Recently, the NL$\sigma$ model attracts interests since the model naturally arises as the effective model of a vortex string in the U($N$) gauge model with $N$  complex scalar fields \cite{Hanany,Auzzi,Eto,Tong,Eto2,Shifman}. 
The NL$\sigma$ model itself has a long history from the proposal as a toy model of the pion \cite{Gellman}. 
The model also arises from the continuum limit of the Heisenberg model. 
This model shares a number of phenomena common with 3+1 dimensional QCD, e.g. asymptotic freedom, dynamical mass generation, confinement, and instantons \cite{Coleman}. 
In this section, we will not trace extensive investigations on the NL$\sigma$ model. 
Instead we will concentrate on the connections with the NJL model. 
Though the NJL model is an interacting fermionic model and the NL$\sigma$ model is the bosonic model with nonlinear constraint (explained below), 
these theories are connected in various ways. 
For instance, the supersymmetric NL$\sigma$ model consists of the NJL part, the NL$\sigma$ part and the fermion-boson interaction term \cite{Witten77}. 
The partition functions of these two theory are known to be connected by a simple formula \cite{Filothodoros}. 
Moreover, the self-consistent equations derived from the NL$\sigma$ model is found to be the same with ones from the NJL model \cite{YoshiiNitta17}.  

\subsection{Model, method, and homogeneous solution}
First, we briefly review the model, method, and homogeneous solutions. 
We consider the O($N$) and the $\mathbb{C}P^{N-1}$ models. 
The Lagrangian of the O($N$) model is given by 
\begin{equation}
L=\int dx \left[\partial_\mu \vec n \cdot \partial^\mu \vec n-\lambda (\vec n \cdot\vec n-r )\right], 
\label{O(N)}
\end{equation}
where $n$ is $N$-components real scalar fields and $\lambda$ is a Lagrange multiplier 
which fixes the normalization condition of the real vector $\vec n$.
Here $r$ is the ``radius" of the vector and is corresponding to the coupling between the bosons. 
The target space of this sigma model is $\mathrm{O}(N)/\mathrm{O}(N-1) \simeq S^{N-1}$.
By replacing the differentiation by a finite differentiation, the Lagrangian becomes
\begin{equation}
L=\sum_i  \frac{\vec n_{i+1}-\vec n_{i}}{a} \cdot \frac{\vec n_{i+1}-\vec n_{i}}{a}
=-\frac{2}{a}\sum_i  \vec n_{i+1}\cdot \vec n_{i}+\mathrm{const}, 
\label{FHeisenberg}
\end{equation}
with the constraint $\sum_i \vec n_i \cdot\vec n_i=r$. Here $a$ is the lattice constant. 
In a 1+1 dimensional lattice, one can separate the system into the even site and the odd site. 
By changing the definition of the field $\vec n_{i\in \mathrm{even}}=-\vec N_i$ and $\vec n_{i\in \mathrm{odd}}=\vec N_i$, 
we obtain 
\begin{equation}
L=\frac{2}{a}\sum_i  \vec N_{i+1}\cdot \vec N_{i}+\mathrm{const}, 
\label{AFHeisenberg}
\end{equation}
where we again have the constraint $\sum_i \vec N_i \cdot\vec N_i=r$. 
The model (\ref{FHeisenberg}) and (\ref{AFHeisenberg}) are the relativistic ferromagnetic and anti-ferromagnetic Heisenberg model, respectively. 
In this model, $\partial_x(\vec n_i \cdot \vec n_i)=0$ and $\partial_t(\vec n_i \cdot \vec n_i)=0$ holds and thus the first order derivative term does not appear. 
More rigorously one need to take into account the Berry phase term in the case of half-integer spins. 
In this article we focus on the case in which the Berry phase term is irrelevant. 

The Lagrangian of the $\mathbb{C}P^{N-1}$ model is given by 
\begin{equation}
L=\int dx \left[D_\mu \vec n^\ast \cdot D^\mu \vec n-\lambda (\vec n^\ast \cdot\vec n-r )\right], 
\label{nls}
\end{equation}
where $D_\mu=\partial_\mu-iA_\mu$ is a covariant derivative and a vector potential $A_\mu$ is an auxiliary field. 
Here $\vec n$ is $N$-components complex scalar fields and $\lambda$ is a Lagrange multiplier 
which fixes the normalization condition of the complex vector $\vec n$ 
\begin{equation}
\vec n^\ast \cdot\vec n=r. 
\end{equation}
Eliminating the auxiliary fields, one finds that the target space of the $\mathbb{C}P^{N-1}$ model is
$\mathbb{C}P^N-1 \simeq \mathrm{SU}(N)/[\mathrm{SU}(N-1)\times \mathrm{U}(1)]$.
Both the O($N$) and the $\mathbb{C}P^{N-1}$ are free boson theories with mass $\lambda$ if $\lambda$ is constant parameter. 
However, we need to consider the stationary condition for $\lambda$ and thus the model has nonlinearity.

In the following we set the vector potential $A_\mu$ to be zero by a gauge transformation since it is not dynamical 
in the leading order of the large-$N$ expansion. 
By this simplification, the O($N$) and $\mathbb{C}P^{N-1}$ model become the same except for the number of components.
In the following, we choose the $\mathbb{C}P^{N-1}$ model for the explanation. 
The vector $\vec n$ can have the vacuum expectation value if the symmetry broken phase is realized. 
By choosing the 0-th component of the field to be classical $n_0=\sigma$, one can obtain the effective Lagrangian for $\lambda$ and $\sigma$ by integrating out the other fields $n_1,\cdots, n_{N-1}$ from the full partition function $\int dn dn^\ast d\lambda \exp(i\int dt L)=\int d\sigma d\lambda \exp(i\int dt L_{\mathrm{eff}})$ \cite{Bolognesi} 
\begin{equation}
L_{\mathrm{eff}}=(N-1)\mathrm{Tr}\log(-\partial_\mu\partial^\mu-\lambda)+ \partial_\mu\sigma\partial^\mu\sigma+\lambda (\sigma^2 -r ), 
\end{equation}
where $N-1$ factor of the first term appears from the integration of $n_1,\cdots, n_{N-1}$ complex fields. 
The stationary condition for $\lambda$ yields 
\begin{equation}
(N-1)\mathrm{Tr}\frac{1}{\partial_\mu\partial^\mu+\lambda}+ \sigma^2-r=0. 
\label{gapeqcpn}
\end{equation}
On the other hand, by assuming the time-independent VEV, the stationary condition for $\sigma$ yields 
\begin{equation}
(-\partial_x^2+\lambda) \sigma=0. 
\label{zeromodecpn}
\end{equation}
In the case of the homogeneous solution $\sigma=\mathrm{const}$, the Eq.\ (\ref{zeromodecpn}) reduces to $\lambda \sigma=0$ and thus we have three possibilities $(\sigma=0, \lambda=0)$, $(\sigma\neq 0, \lambda=0)$, or $(\sigma=0, \lambda\neq 0)$. 
In the infinite system, the first term of the Eq.\ (\ref{gapeqcpn}) has the infrared divergence for $\lambda$ and thus the third possibility only survives. 
Thus, all $N-1$ bosons have the dynamical mass $\lambda\neq 0$ in the infinite system. 
This is consistent with the Coleman-Mermin-Wagner theorem \cite{Coleman2, Mermin} which inhibits the appearance of the NG bosons in 1+1 dimensions. 

It should be noted about the difference between the NL$\sigma$ model and NJL model. 
In the NL$\sigma$ma model, the massless modes (NG boson) appear in the symmetry broken phase where the VEV is nonzero ($\sigma\neq 0$ solution), if such a solution exists. 
On the other hand, the energy gap appears in the symmetric phase. 
This is completely opposite behavior with the NJL model where the energy gap appears in the broken phase and the massless mode appears in the symmetric phase. 
It should also be noted that the above method is also valid if we replace the complex fields to the real fields, namely $\mathbb{C}P^{N-1}$ model to the O($N$) model. 

\subsection{Inhomogeneous solutions}
In the previous subsections, we have assumed condensations to be homogeneous. 
In order to extend the method presented above to inhomogeneous condensations, we need to consider the spatial dependences of $\lambda$ and $\sigma$. 
The main modification is that the trace in Eq.\ (\ref{gapeqcpn}). 
In the case of the homogenous condesation, the trace is taken with plane wave solutions. 
On the other hand, the eigenmodes are no longer plane waves and thus we 
need to consider the following eigenvalue equation \cite{Bolognesi}
\begin{equation}
[-\partial_x^2+\lambda(x)]f_n(x)=\omega_n^2 f_n(x), 
\label{eigencpn}
\end{equation}
where $\omega_n$ and $f_n(x)$ are the $n$-th eigenvalue and eigenstate of the operator $-\partial_x^2+\lambda(x)$, respectively. 
By using this eigenvalue, Eq.\ (\ref{gapeqcpn}) can be rewritten as 
\begin{equation}
\frac{N-1}{2}\sum_n \frac{f_n(x)^2}{\omega_n}+ \sigma(x)^2-r=0. 
\label{gapeqcpn2}
\end{equation}
Thus, we have the three equations (\ref{zeromodecpn}), (\ref{eigencpn}), and (\ref{gapeqcpn2}) to be solved self-consistently. 

The three self-consistent equation in NL$\sigma$ models can be connected with the two self-consistent equations (BdG equation and gap equation ) for the NJL model by introducing the auxiliary field $\Xi$ satisfying the Riccati equation \cite{YoshiiNitta17}
\begin{equation}
\Xi^2+\partial_x \Xi=\lambda. 
\end{equation}
The correspondence between the NJL and the NL$\sigma$ model can be seen by identifying $\Xi$ and $\Delta$. 
The equation for $\sigma$ (\ref{zeromodecpn}) can be rewritten as $(\partial_x+\Xi)(-\partial_x+\Xi)\sigma=0$ and is the same with the zero mode equation for the fermionic system. 
The eigenvalue equation (\ref{eigencpn}) is nothing but the Klein-Gordon-like equation (\ref{KGeq}) obtained from the BdG equation (\ref{BdGChiral}). 
Thus, the corresponding BdG-like equation can be constructed by a simple guess: 
\begin{equation}
\left(
\begin{array}{cc}
0 & \partial_x+\Xi \\
-\partial_x+\Xi & 0
\end{array}
\right)
\left(
\begin{array}{cc}
u_n \\
v_n
\end{array}
\right)
=E\left(
\begin{array}{cc}
u_n \\
v_n 
\end{array}
\right).
\label{BdGCPN}
\end{equation}
Differentiating the third equation (\ref{gapeqcpn2}) by $x$ and using the other equations, one can show that the differentiated (\ref{gapeqcpn2}) is equivalent to the gap equation in the NJL model 
\begin{equation}
\Xi=\frac{N}{2r}\sum_n u_nv_n,  
\end{equation}
where $u_n$, $v_n$ are the upper and lower component of the $n$-th eigenspinor for the BdG equation. 
We note that the summation of $n$ is taken for positive eigenvalues, whereas the summation is taken for negative eigenvalues for the NJL model. 
This difference changes the overall sign in the right hand side of the gap equation. 
It is obvious to see that the role of the coupling $g$ in the NJL model is played by $1/r$ in the NL$\sigma$ model. 
According to this, it is also known that the coupling constant $g_{\mathrm{YM}}$ in the Yang-Mills theory is connected with the radius as $r=4\pi/g_{\mathrm{YM}}^2$ when the $\mathbb{C}P^{N-1}$ model is realized as the effective theory of a vortex line in the U($N$) gauge theory. 

It should be noted that we need to check the self-consistency by plugging the solutions into (\ref{gapeqcpn2}) since the above procedure only guarantees the differentiated version of Eq.\ (\ref{gapeqcpn2}) which does not contain the information of the integral constant.  
This last step gives a significant difference between the NJL and NL$\sigma$ models. 
It is easily checked that the Higgs phase $\sigma\neq 0$ in the NL$\sigma$ model corresponds to the normal phase $\Delta=0$ in the NJL model. 
The former solution is inhibited and the latter is allowed though the normal phase is excited state in the infinite system.  
The absence of the Higgs phase indeed comes from the fact that there is no $\sigma\neq 0$ solution 
which satisfies Eq.~(\ref{gapeqcpn}) although the differentiated one can be satisfied. 

\begin{figure}
\includegraphics[width=10 cm]{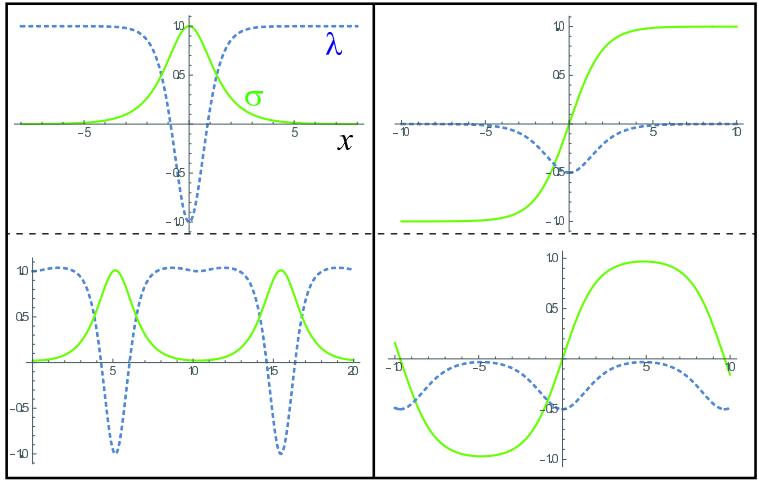}
\caption{Inhomogeneous solutions for the NL$\sigma$ model. 
The solid lines and the dashed lines correspond to the Higgs field configuration $\sigma$ and the gap function $\lambda$, respectively. 
In the all figures, we set $m=1$ and normalize the Higgs fields $\sigma$ as it fits to the figure. 
The upper left figure corresponds to the Higgs soliton, where the Higgs field (magnetization) has the nonzero value around $x=0$. 
The bottom left figure is the Higgs soliton lattice solution with $\nu=0.9$. 
In the upper right figure, the confining soliton solution is shown.  
The bottom right figure is the confining soliton lattice solution for $\nu=0.99$. }
\label{inhomoCPN}
\end{figure}

The NJL/NL$\sigma$ correspondence gives various new inhomogeneous solutions in the NL$\sigma$ model by translating the inhomogeneous solutions known in the NJL model. 
The kink solution (\ref{kinksol}) in the NJL model corresponds to the following Higgs soliton solution 
\begin{equation}
\lambda=m^2\left(1-\frac{2}{\cosh^2 mx}\right),\quad  
\sigma=\frac{A_{cs}}{\cosh mx}, 
\end{equation}
where $A_{cs}=m\sqrt{N\sum_n\omega_n^{-3}}$ is the normalization factor. 
In Fig.\ \ref{inhomoCPN} (upper left figure), we plot the gap function and the vacuum expectation value of the Higgs soliton solution. 
For the case of the Higgs soliton solution, the symmetry is broken in the localized region around $x=0$, 
whereas the dual solution in the NJL model (kink solution) locally restores the symmetry around $x=0$. 
Since this solution does not posses the long-range order, even the VEV is nonzero, 
it is consistent with the Colman-Mermin-Wagner theorem \cite{Coleman2, Mermin}. 
The energy difference from the homogeneous confining solution can be calculated as \cite{YoshiiNitta17}
\begin{equation}
\Delta E=E_{\mathrm{HS}}-E_{\mathrm{Conf}}=4rm, 
\label{EHiggsSol}
\end{equation} 
where $E_{\mathrm{HS}}$ and $E_{\mathrm{Conf}}$ are the total energy for the Higgs soliton solution and the confining solution, respectively. 

The generalization to a multi-soliton solution is possible by mapping the kink-crystal solution in the NJL model (\ref{kinkcrystalsol}) 
following the NJL/NL$\sigma$ prescription. 
In such a case, the following Higgs soliton lattice solution is obtained: 
\begin{equation}
\lambda=m^2\left[\mathrm{sn}(mx,\nu)^2+\mathrm{cn}(mx,\nu)\mathrm{dn}(mx,\nu) \right],\quad 
\sigma=A_{csl}\left[\frac{-\sqrt{\nu}\mathrm{cn}(mx,\nu)+\mathrm{dn}(mx,\nu)}{1-\sqrt{\nu}}\right]^{\pm 1/\sqrt{\nu}}, 
\end{equation}
where $\mathrm{cn}, \mathrm{dn}$ are the Jacobi's elliptic functions. 
For this solution (lower left figure of Fig.\ \ref{inhomoCPN}), the symmetry is locally broken and the broken regions are periodically aligned. 
The more general solutions belonging to the higher rank of the AKNS hierarchy in the NJL model can also be mapped to the solutions in the NL$\sigma$ model (for more details, see Ref.~\cite{YN2018-1}). 
The method is also applicable to the case of the supersymmetric $\mathbb{C}P^{N-1}$ model and the inhomogeneous solutions are also found there \cite{Gorsky}.

The confining soliton solution could be chosen in the presence of the ``magnetic"-impurity or the local ``magnetic"-field 
inducing the term $\tilde B\cdot n_0(x) \delta(x)$ to the Hamiltonian, similar to the kink solution in the polyacetylene. 
In the presence of the periodic structure of the ``magnetic" impurity, the Higgs soliton lattice solution is expected as a candidate of the ground state. 
Thus, it should be expected if the ``magnetic"-field $\tilde B$ is much larger than $4rm$ (see Eq.~{EHiggsSol}), the appearance of the soliton (local magnetization) is preferred \cite{footnote5}.

\subsection{Finite system}
For the case of the finite systems, the boundary condition, finite size effect, and/or topology can induce the various effects. 
The numerical simulation was done in Refs.\ \cite{Bolognesi,Betti} for the Dirichlet boundary condition and the inhomogeneous solution was found. 
The absence of the homogeneous solution for such system is also proven. 
The Casimir force for the finite interval was investigated in Ref.\ \cite{FNTYCPN, Chernodub1, Betti}. 
More recently the Casimir force for the Yang-Mills theory was also investigated \cite{Chernodub2}.
This was also proved by the analytical approach \cite{FNTYCPN}. 
In 2+1 dimensions, a disc system with the Dirichlet boundary condition at the edge (a circle) was investigated and the absence of the homogeneous solution was also shown \cite{Pikalov}. 
In the case of a ring geometry, the phase transition between the confining phase and the Higgs phase is reported 
where the IR divergence which inhibits the Higgs phase was removed by a natural IR cutoff corresponding to the inverse of the system size \cite{Monin}. 
The mixed boundary condition such that a half ($N$/2) of the components obey the Dirichlet boundary condition and the rests obey the Neumann condition was investigated \cite{Milekhin,Pavshinkin}. 
In this system, the homogeneous solution is allowed and the Higgs phase appears because of the IR cutoff induced by the system size. 
This allowance of the Higgs phase makes the phase structure of the system to be richer, 
where for instance a confining soliton in the Higgs background can exist \cite{YN2018-1}. 

The inhomogeneous solution in the Higgs background 
can be made from the partially-symmetry-broken solution $\Delta=m/\sinh mx$ (Eq.\ (\ref{cs}) with $\nu\to 1$) in the NJL model \cite{YN2018-1}: 
\begin{equation}
\lambda=-\frac{m^2}{2\cosh^2 mx/2},\quad  
\sigma=A\tanh \frac{mx}{2}.
\label{HK}
\end{equation}
This confining soliton solution corresponds to a partially-symmetry-restored solution. 
The multiple soliton solutions can also be constructed by using the general solution (\ref{cs}): 
\begin{equation}
\lambda=\frac{\mathrm{cn}(mx,\nu)^2-\mathrm{dn}(mx,\nu)}{\mathrm{sn}(mx,\nu)^2},\quad 
\sigma=A\frac{1-\mathrm{dn}(mx,\nu)}{\mathrm{sn}(mx,\nu)}.
\label{HKL}
\end{equation}
We plot the confining soliton solution and the confining soliton lattice solution in the upper right and lower right figures in Fig.\ \ref{inhomoCPN}, respectively.  
We can evaluate the difference between the total energy of the constant Higgs solution $E_{Higgs}$ 
and the Higgs kink solution $E_{HK}$ as 
\begin{equation}
\Delta E=E_{HK}-E_{Higgs}=rm. 
\end{equation} 
Similar to the energy difference between the confining solution and the Higgs soliton solution (\ref{EHiggsSol}),  
the energy to make the soliton to the homogeneous background is $\sim rm$. 

This solution might be chosen in the presence of the constraint that the orientation of VEV at the left edge and the right edge is opposite as a boundary condition. 
In the case that the orientation of VEV is not parallel, the phase-winding solution similar to the FF state in the NJL model could be stabilized. 
According to that, the solution with two component Higgs fields is obtained where the winding of VEV occurs \cite{YN2018-2}. 

\subsection{NJL/$\mathbb{C}P^{N-1}$ correspondence in 2+1 dimension}\label{3DNJLNLS}
In this subsection, we briefly review the other approach to a NJL/$\mathbb{C}P^{N-1}$ correspondence in higher dimensions \cite{Filothodoros}. 
The main idea is as follows. 
If one changes the boundary condition in the imaginary time direction, the statistical transmutation may take place.  
This intuition comes from the fact that the fermionic fields $\psi$ obey the anti-periodic boundary condition 
\begin{equation}
\psi(\tau)=-\psi(\tau+\beta), 
\end{equation}
whereas the bosonic fields $\phi$ obey the periodic boundary condition 
\begin{equation}
\phi(\tau)=\phi(\tau+\beta), 
\end{equation}
in the  imaginary time direction. 
The change of the boundary condition is realizable by the following gauge transformation 
\begin{equation}
f(\tau)\to\psi^\prime(\tau)=e^{i\int_0^\tau d\tau^\prime a_0(\tau^\prime)}f(\tau), 
\end{equation}
where $f$ is either $\psi$ or $\phi$, $a_0$ is the 0th component of the gauge field.  
If $\int_\tau^{\tau+\beta} d\tau^\prime a_0(\tau^\prime)=\pi$ (mod $2\pi$), 
the periodic (anti-periodic) boundary condition is mapped anti-periodic (periodic) boundary condition. 

In the presence of the 0th component of the gauge field, the derivative with respect to the imaginary time $\partial_\tau$ becomes $\partial_\tau-ia_0$. 
This gauge field is naturally introduced by imposing the canonical condition 
\begin{equation}
Z(\beta,N)=\mathrm{Tr}\left[\delta(N-\hat N)e^{-\beta\hat H}\right]
=\int_0^{2\pi}\frac{d\theta}{2\pi} e^{i\theta N}\mathrm{Tr}\left[e^{-\beta\hat H-i\theta \hat N}\right], 
\end{equation}
where $\hat N$ is the number operator and $N$ is the eigenvalue of that. 
The eigenvalue of $\hat N$ is positive integer and thus the Fourier conjugate value $\theta$ is periodic ($\theta\in [0,2\pi]$). 
The term $\mathrm{Tr}\left[e^{-\beta\hat H-i\theta \hat N}\right]$ in the above equation can be seen as the grand canonical partition function with the imaginary chemical potential $-i\theta/\beta$. 
It is found that the NJL model with the imaginary chemical potential $\mu=-ia_0$ with $a_0=0$ yields the free energy in the unit volume ($[-\beta^{-1}\ln Z]/V$) 
measured from that in the absolute zero temperature as
\begin{equation}
f_F^0(\beta)-f_F^0(\infty)=-N\frac{3}{2}\frac{\zeta(3)}{2\pi\beta^3}, 
\end{equation} 
whereas the case of the $\alpha=\pi/\beta$, it becomes 
\begin{equation}
f_F^{\pi}(\beta)-f_F^{\pi}(\infty)=N\frac{8}{5}\frac{\zeta(3)}{2\pi\beta^3}, 
\end{equation} 
where $\zeta$ is the Riemann's zeta function. 
On the other hand, in the case of the $\mathbb{C}P^{N-1}$ model with the imaginary chemical potential $\mu=-ia_0$ it becomes 
\begin{equation}
f_B^0(\beta)-f_B^0(\infty)=-N\frac{8}{5}\frac{\zeta(3)}{2\pi\beta^3}, 
\end{equation} 
whereas the case of the $\alpha=\pi/\beta$, it becomes 
\begin{equation}
f_B^{\pi}(\beta)-f_B^{\pi}(\infty)=N\frac{3}{2}\frac{\zeta(3)}{2\pi\beta^3}. 
\end{equation} 
The above results imply that the NJL/$\mathbb{C}P^{N-1}$ correspondence shown in the 1+1 dimensions may not be accidental coincidence.

\section{Summary}
In this review, we have presented the connections between the condensed matter systems and the NJL and nonlinear $\sigma$ models. 
As examples we have considered superconducting phenomena in metallic systems and a charge density wave in polyacetylene systems. 
In the view of spontaneous symmetry breaking, those phenomena can be understood in the same theoretical framework. 
Furthermore, we have shown that the various method invented in high energy physics can be applicable to condensed matter systems and vice versa, 
especially concerning to inhomogeneous condensations. 

Because of the limitation of the space, several important issues are lacking or just superficially mentioned. 
For instance, the dynamics of the condensation is quite important in order to understand responses to external fields and the transport phenomena. 
The other important issue should be finite temperature and/or finite density effects. 

The inhomogeneous solutions stabilized by external conditions and/or the external fields have been discussed. 
In low dimensions, such inhomogeneous solutions can possibly induce fractional charges and/or spins and those solutions represent condensates in the presence of defects. 
The external fields, for example the magnetic field on a supeconductor, can also induce inhomogeneous solutions, 
in which the inhomogeneity is determined by the strength of external fields. 
In the case of superconductors, a rich phase diagram with the first order transition lines is obtained. 
The inhomogeneity generation for the polyacetylen system by the external field might be interesting problem. 

It has been also shown that the NJL model shares a mathematical structure with NL$\sigma$ models. 
Though a deep understanding is still lacking, 
a duality between the partition functions in 2+1 dimensional systems strongly suggests an underlying connection. 
The direct mapping between the two theories in 1+1 dimensions has been found and it yields various inhomogeneous solutions in NL$\sigma$ models. 
The dual structure of solutions is shown; 
the symmetry-broken(restored)-phase in the NJL model corresponds to the restored(broken)-phase in the NL$\sigma$ models. 
This duality is also generalized to partially broken (restored) solutions. 
The generalization to finite temperature and/or density case should be important and we leave it for a future problem.

\acknowledgments{We would like to thank Antonino Flachi, Hisao Hayakawa, Giacomo Marmorini, Satoshi Takada, Daisuke Takahashi, and Shunji Tsuchiya 
for collaborations of the original works on which this review article is based.. This work is supported by the Ministry of Education, Culture, Sports, Science (MEXT)-Supported Program for the Strategic Research Foundation at Private Universities eTopological Sciencef (Grant No. S1511006). The work of M.\ N.\ is also supported in part by the Japan Society for the Promotion of Science (JSPS) Grant-in-Aid for Scientific Research (KAKENHI Grant No. 16H03984 and No. 18H01217) and a Grant-in-Aid for Scientific Research on Innovative Areas ``Topological Materials Science" (KAKENHI Grant No. 15H05855).}


\begin{thebibliography}{120}
\bibitem{footnote1}
In this article, we focus on the O($N$) sigma model and the $\mathbb{C}P^{N-1}$ model. Thus the NL$\sigma$ model in the following always means either or both of them.
\bibitem{footnote2}
For the superconductor, we should use the terminology ``1+1 dimensional NJL model" or ``chiral GN model", more precisely. Similarly, we should use ``$\mathbb{Z}_2$ NJL model" instead of ``NJL model" or simply ``GN model" for the system in which the condensate only takes a real value as a polyacetylene.
\bibitem{footnote3}
Conventionally, $A_k$ and $A_k\alpha_k$ are written as $u_k$ and $v_k$, respectively.
\bibitem{footnote4}
We note that there are other possibilities in the presence of the population imbalance. 
For instance, the gapless state is proposed by pairing the Cooper pair as the total moment to be zero. 
In the case, the superconducting resonant state is formed around the fermi surface of the minority spins, 
which is below the fermi surface of the majority spins \cite{Sarma, Wilczek}. 
Basically this phase, the so-called Sarma phase, is known to be unstable (indeed it gives maximum of the thermodynamic potential) in usual BCS model.
\bibitem{footnote5}
There is a claim that the inhomogeneous solutions are energetically favorable if the anomalous contribution is properly considered \cite{Gorsky}.
\end{thebibliography}
\end{document}